\documentclass[a4paper, amsfonts, amssymb, amsmath, reprint, showkeys, nofootinbib, twoside]{revtex4-1}
\usepackage[english]{babel}
\usepackage[utf8]{inputenc}
\usepackage[colorinlistoftodos, color=green!40, prependcaption]{todonotes}
\usepackage[pdftex, pdftitle={Article}, pdfauthor={Author}]{hyperref} 

\usepackage{amsthm}
\usepackage{mathtools}
\usepackage{xcolor}
\usepackage{graphicx}
\usepackage[left=13mm,right=13mm,top=25mm,columnsep=15pt]{geometry} 
\usepackage{adjustbox}
\usepackage{placeins}
\usepackage[T1]{fontenc}
\usepackage{lipsum}
\usepackage{csquotes}

\begin{document}

\title{Breaking the Brownian Barrier: Models and Manifestations of Molecular Diffusion in Complex Fluids}

\author{Harish Srinivasan}
	\email[]{harishs@barc.gov.in}
	\affiliation{Solid State Physics Division, Bhabha Atomic Research Centre, Mumbai 400085}
	\affiliation{Homi Bhabha National Institute, Mumbai 400094}

\author{V. K. Sharma}
	\email[]{sharmavk@barc.gov.in}
	\affiliation{Solid State Physics Division, Bhabha Atomic Research Centre, Mumbai 400085}
	\affiliation{Homi Bhabha National Institute, Mumbai 400094}
	
\author{S. Mitra}
	\affiliation{Solid State Physics Division, Bhabha Atomic Research Centre, Mumbai 400085}
	\affiliation{Homi Bhabha National Institute, Mumbai 400094}
	
\begin{abstract}
Over a century ago, Einstein formulated a precise mathematical model for describing Brownian motion. While this model adequately explains the diffusion of micron-sized particles in fluids, its limitations become apparent when applied to molecular self-diffusion in fluids. The foundational principles of Gaussianity and Markovianity, central to the Brownian diffusion paradigm, are insufficient for describing molecular diffusion, particularly in complex fluids characterized by intricate intermolecular interactions and hindered relaxation processes. This perspective delves into the nuanced behavior observed in diverse complex fluids, including molecular self-assembly, deep eutectic solvents, and ionic liquids, with a specific focus on modeling self-diffusion within these media. We explore the potential of extending diffusion models to incorporate non-Gaussian and non-Markovian effects by augmenting the Brownian model using non-local diffusion equations. Further, we validate the applicability of these models by utilizing them to describe results from quasielastic neutron scattering and MD simulations.
\end{abstract}

\maketitle


\section{Historical overview of Brownian diffusion}
The notion of random motion has been a topic of great philosophical and scientific discourse throughout the history of humanity. As one of the earliest proponents of this idea, a Roman poet and philosopher Lucretius in his scientific poem "On the nature of Things" (c. 60 BC) wrote the following about behaviour of dust particles:
\begin{quote}
"And thus they flit around in all directions,\\
At random all and every way, and fill\\
The hidden nooks of things with a seething whirl."    
\end{quote}
An idea he proposed to reinforce the atomistic models in Roman philosophy bears a gross resemblance to the nature of the random motion of particles suspended in a fluid. However, the first true discovery of random motion of particles in a fluid is credited to the botanist Robert Brown. In 1827, during a study on the suspension of pollen grains in water, he observed numerous particles ejected by pollen grains performing zig-zag motion. Through a series of observations on different materials, Brown noted that this motion was not related to life, but rather a physical phenomenon which had no biological origin. Many years later, Einstein deduced a mathematical model of Brownian motion from statistical physics unawares of the observation of Brown.  

\subsection{Einstein's model}
In his groundbreaking 1905 paper \cite{Einstein_1905}, Einstein built upon the progress in molecular kinetic theory, primarily pioneered by the efforts of Boltzmann, Maxwell, and Gibbs. By employing these fundamental principles, Einstein addressed the phenomenon of Brownian motion, elucidating the molecular or atomic underpinnings of the persistent and irregular motion displayed by microscopic particles suspended in a stationary fluid.

Owing to the probabilistic nature of particle dynamics, a probability distribution function (PDF), $P(x,t)$, characterizing the probability of finding the particle at some position $x$ at a time $t$, is considered along with certain initial conditions $P(x,0)$. In Einstein's model, the Brownian motion of particles is propagated by making small displacements, of length $\Delta$. These displacements essentially occur due to random molecular collisions on the particle and are characterized by a probability distribution, $f(\Delta)$. An equation governing the time-evolution of the PDF $P(x,t + \tau)$ can the be given as \cite{Einstein_1905},
\begin{equation}
    \label{Einstein}
    P(x, t + \tau) = \int\displaylimits\displaylimits_{-\infty}^{\infty} d \Delta \; f(\Delta) P(x-\Delta,t)
\end{equation}
The above equation quantitatively relates the probability of detecting particles at position $x$ at time $t + \tau$ to two factors: (i) the likelihood of encountering the particle at $x-\Delta$ at time $t$ and (ii) the probability of the particle undergoing a displacement of length $\Delta$. It is crucial to emphasize several key assumptions that form the foundation of the aforementioned equation. Foremost among these assumptions is that it considers the occurrence of a single displacement within the time interval $\tau$. Additionally, the equation assumes that the probability of the jump's magnitude remains unaffected by the displacements at an earlier time. These assumptions align naturally with the meaningful scales inherent to the time interval $\tau$. This interval is considerably small when compared to the macroscopic time frame of experimental measurements, a factor that allows it to be treated as a single displacement. However, it retains a reasonably large magnitude when compared to average time-interval between molecular collisions, thereby maintaining the assumption of independence between consecutive jumps. In order to obtain a rate equation governing the probability density $P(x,t)$, we consider the Taylor expansion of \eqref{Einstein} with respect to $\tau$ on the RHS and $\Delta$ on the LHS, yielding,
\begin{equation}
\begin{split}
    \label{Einstein2}
    P(x,t) + \tau \frac{\partial P(x,t)}{\partial t} + O(\tau^2) \\= \int\displaylimits_{-\infty}^{\infty} d\Delta f(\Delta) \left[ P(x,t) - \Delta \frac{\partial P}{\partial x} + \frac{\Delta^2}{2} \frac{\partial^2 P}{\partial x^2} + O(\Delta^3) \right]
\end{split}
\end{equation}
The first term on the sides cancel out, owing to the normalization of $f(\Delta)$. Further, considering that the suspended particles experience isotropic collisions, enforces that the mean jump-length is zero ($\left< \Delta \right> = 0$). Therefore we have,
\begin{equation}
    \label{Einstein3}
    \frac{\partial P(x,t)}{\partial t} = \frac{\left< \Delta^2 \right>}{2 \tau} \frac{\partial^2 P(x,t)}{\partial x^2}
\end{equation}
where we have retained terms of only lowest order with non-zero contribution in the Taylor expansion. This truncation is equivalent to the physical assumption that the observation length $x \gg \Delta$ and time scales $t \gg \tau$. It is easy to recognize the structure of eq. \eqref{Einstein3} as the diffusion equation, with the diffusivity, $D$ given by $\left< \Delta^2 \right>/(2\tau)$. The solutions to eq. \eqref{Einstein3} is given by a standard Gaussian distribution, for an initial condition $P(x,0) = \delta(x)$, as, \cite{Einstein_1905, zwanzig2001nonequilibrium}
\begin{equation}
    \label{Einsteinsoln}
    P(x,t) = \frac{1}{\sqrt{4\pi D t}} e^{-\frac{x^2}{4 D t}}
\end{equation}
Further, an immediate consequence \eqref{Einsteinsoln} is the nature of average particle displacements. Due to the isotropic nature of the motion, mean displacement $\left< x \right>$ is zero. However, the mean-squared displacement (MSD), $\left< x^2 \right> = 2 D t$, exhibits a linear relationship with time. The square-root time-dependence of displacement is a distinctive feature of Brownian motion.  

\subsection{Langevin's model - treatment of velocity}

While Einstein's model provided a breakthrough in understanding of molecular kinetic theory it primarily lacked the description of velocity of the Brownian particles in the system. In 1908, Langevin provided a model \cite{Langevin_1908} to describe the velocity of the diffusing particle. At its core, Langevin's model aims to describe how a single particle moves through a viscous medium (bath), like a small particle suspended in a liquid. The equation governing the velocity of the particle is given by,
\begin{equation}
    \label{Langevin}
   m \frac{d v}{dt} = - \gamma v + \xi(t)  
\end{equation}
The equation describes the motion of a particle in a fluid, influenced by viscous drag and random thermal fluctuations.
The first term represents the viscous drag characterized by frictional constant, $\gamma$. The second term, $\xi(t)$ is the random force experienced by the Brownian particles due to collision atoms/molecules of the fluid. The random force, $\xi(t)$ is referred to as a stochastic process which represents a random variable that varies in time \cite{gardiner2009} and is defined through it's statistical measures or ensemble averages (denoted by $\langle\rangle$). The stochastic process employed in the Langevin equation \eqref{Langevin} is called the Gaussian white noise. Notably, it has zero average, $\left<\xi(t)\right>=0$, with an underlying Gaussian distribution, reflecting the isotropic nature of collisions on the Brownian particle. Secondly, the variance of $\xi(t)$ is given by a delta correlation, indicating that no two collisions happening at two different instants of time are correlated. Mathematically, this is given as, $\left< \xi(t) \xi(t') \right> = 2 \gamma k_B T \delta(t-t')$, where $T$ is the temperature of the bath and $k_B$ is the Boltzmann constant. This relationship between the variance of $\xi(t)$ and frictional constant $\gamma$ is fixed by the conditions of thermal equilibrium at a temperature $T$ and is referred to as the fluctuation dissipation theorem. 

The Langevin's model treats particle velocity as a stochastic process, driven by another stochastic process, $\xi(t)$, which is a Gaussian white noise. Notably, the velocity also exhibits a Gaussian distribution owing to the underlying Gaussian nature of the white noise. Moreover, the velocity of particles exhibits Markovian behavior, meaning it has no history dependence and only depends on its current state. This Markovian property arises from the delta-correlation characteristic of the white noise term in the Langevin equation. Physically, this indicates that the size of Brownian particle is sufficiently large, such that subsequent collisions are completely uncorrelated. Later, it shall be discussed how this model breaks down, when molecular collisions become correlated and Markovianity is no more preserved.

The behaviour of velocity of Brownian particle can be obtained from eq. \eqref{Langevin} by calculating the velocity autocorrelation function (VACF), $C_v(t)$, given by,
\begin{equation}
    \label{Langevin_VACF}
    C_v(t) = \left< v(t) v(0) \right> = \frac{k_B T}{m} \exp\left[-\gamma t/m  \right]
\end{equation}
VACF decays exponentially with the decay rate being dictated by two physical attributes, frictional constant, $\gamma$ and mass of the particle, $m$. The decay of the VACF is enhanced for higher frictional constant and lower mass of the particle. In the case, when the frictional force is higher in the liquid, VACF decays faster owing to larger number of collisions by molecules of the fluid. On the other hand, when particle is more massive, it is less likely to change it's initial velocity due to stronger inertial effects. The VACF can also be used to calculate the mean-squared displacement (MSD) of the diffusing particle from the Green-Kubo relationship
\begin{equation}
    \label{green-kubo}
    \begin{split}
    \left< x^2(t) \right> = 2 \int\displaylimits_0^t \left(t-t'\right)\left< v(t) v(t') \right> dt' \\
    = 2 \frac{k_B T}{\gamma} \left[ t + \frac{m}{\gamma}\left(e^{-\gamma t/m} -1 \right) \right]
    \end{split}
\end{equation}
At sufficiently long times (particularly for $t \gg m/\gamma$), it is evident that the MSD attains the linear time dependence, $\left< x^2(t)\right> \sim 2 \left(k_B T/\gamma\right) t $, which resembles the behaviour of MSD from the Einstein's formalism based on diffusion equation \eqref{Einstein3} considering $D = k_B T/\gamma$.

Diffusion constant, $D$ is a crucial physical parameter in the diffusion process. In the Einstein's prescription, we found it was linked to the ratio between mean-squared displacement and waiting time between them. On the other hand, in the case of Langevin's model, it is linked to the frictional constant, $\gamma$. Through Stokes equation, the frictional force on a particle is the linked to viscosity of the liquid,  $\eta$, according $\gamma = 6\pi \eta r$, where $r$ is the radius of the diffusing particle. This provides the Stokes-Einstein relationship - link between diffusivity, $D$ and viscosity $\eta$,
\begin{equation}
    \label{SE}
    D = \frac{k_B T}{6\pi \eta r}
\end{equation}

\subsection{Equivalence of Einstein and Langevin models}
While it is clear that at the long-time limit, both Einstein and Langevin models accurately reproduce the linear time-dependence of MSD, and the Gaussian nature of probability distributions. Both the underlying physical motivation and mathematical structure of these models are fundamentally different and complementary in approach. The Einstein model directly deals with probability density of the position of the diffusing particle and provides a basis for Fokker-Planck equation (FPE). On the other hand, Langevin model essentially sets the stage for directly solving differentials involving stochastic processes leading to the formulation of stochastic differential equations (SDEs). It is possible to obtain a one-one correspondence between FPEs and SDEs under certain circumstances. We will explore this in trying to establish the link between these two seminal models of diffusion by considering the long-time limiting behaviour of the Langevin equation \eqref{Langevin}. This limit can also be achieved equivalently by considering a massive particle, referred to as the inertial limit. In either case the contribution of the acceleration to the Langevin equation (eq. \eqref{Langevin}) is ignored, leading to,
\begin{equation}
    \label{Langevin_inertial}
    \frac{d x}{dt} = \sqrt{2D} \chi(t)
\end{equation}
where $\chi(t) = \xi(t)/\left(\sqrt{2\gamma k_B T}\right)$, is a dimensionless Gaussian white noise of unit variance. In this limit, the velocity of the particle behaves as a Gaussian white noise and the position of the particle is obtained as an integral over the white noise,
\begin{equation}
    \label{Wiener_process}
    x(t) = \sqrt{2D} \int\displaylimits_0^t dt' \; \chi(t')
\end{equation}
The integral over the dimensionless white-noise, $\chi(t)$ is also called as Wiener's process ($W(t)$) and holds a special position in calculus of stochastic processes \cite{gardiner2009}. If $P(x,t)$ is the probability density associated to the stochastic process $x(t)$ in eq. \eqref{Wiener_process}, the FPE equivalent equation for the SDE in eq. \eqref{Langevin_inertial} is \cite{gardiner2009},

\begin{equation}
    \label{diffusion}
    \frac{\partial P(x,t)}{\partial t} = D \frac{\partial^2 P(x,t)}{\partial x^2}
\end{equation}
Clearly, this equation resembles the equation (eq. \eqref{Einstein3}) derived in the formalism provided by Einstein, when treating Brownian motion using a random walk model. The relationship between eq. \eqref{Langevin_inertial} and \eqref{diffusion} is a canonical example of FPE $\leftrightarrow$ SDE correspondence.

Two essential features emerge in the solution of Brownian motion in either of these frameworks. Firstly, the mean-squared displacement, $\left< x^2(t) \right>$ varies linearly with respect to time, $t$, and this behaviour is commonly referred to as \textbf{Fickianity}. Secondly, the \textbf{Gaussianity} exhibited by distribution of displacement which arises from solutions of eq. \eqref{diffusion}. As a concluding remark, we note the two fundamental physical assumptions that has been used in the description of Brownian motion model.
\begin{itemize}
    \item \textbf{Markovianity} refers to lack of history dependence in the diffusion process. This assumption is grounded in the physical concept that, for particles with size or mass greater than the molecules in the fluid, collisions upon the particle are uncorrelated.
    \item \textbf{Gaussianity} pertains to the character of displacement distribution being Gaussian in nature. This behaviour is achieved as a limit of large number of small displacements, as a manifestation of central limit theorem. 
\end{itemize}

As we shall see in the subsequent sections, both these assumptions tend to breakdown when considering the description of diffusion of molecules within complex fluids. Under such circumstances, we shall employ more general models which can describe the diffusion based on mathematically consistent models. Before, venturing into generalized extensions beyond the standard Brownian model, we briefly explore the important correlation functions that serve as a window to discern the nature of diffusion in these complex media.

\subsection{Time-correlation functions}
The treatment of displacement, $x(t)$, and velocity $v(t)$, of Brownian particle as stochastic processes is central to most prescriptions of diffusion phenomena. The properties of these stochastic processes are reflected in the behaviour of time-correlation functions of these quantities. The velocity autocorrelation function (VACF) and mean-squared displacement (MSD) discussed in the preceding sections are some of the key estimators.

While the complete PDF, $P(x,t)$ contains information about the nature of diffusion process, it's often not analytically tractable. More often, it is possible to solve for the PDF in the Fourier space. The spatial Fourier transform of the PDF is referred to as the self-intermediate scattering function (SISF), and can be calculated using,
\begin{equation}
    \label{iqt-defn}
    I(k,t) = \int\displaylimits_{-\infty}^\infty dx\;e^{-ikx} P(x,t) = \left< e^{-ik\left[ (x(t) - x(0) \right]} \right>
\end{equation}
SISF holds a special place in the theory of stochastic processes, as they also serves as the moment generating function of $P(x,t)$. Therefore, we have
\begin{equation}
    \label{moments}
    \begin{split}
        \left< x^{2n}(t) \right> = (-1)^n \lim_{k\rightarrow 0}\frac{\partial^{2n} I(k,t)}{\partial k^{2n}} 
    \end{split}
\end{equation}
where only the even moments of displacement are represented, since the odd moments vanish due to the symmetry of the diffusion problem.
The non-Gaussian parameter is another important correlation function which can be a useful quantitative estimator of deviation from Gaussian distribution. The non-Gaussian parameter is defined as a ratio of the second and fourth moments of the position of the particle according to,
\begin{equation}
    \label{NG 1D}
    \alpha_2 (t) = \frac{\left< x^4(t) \right>}{3\left[\left< x^2(t) \right>\right]^2} - 1
\end{equation}
wherein $\alpha_2(t) = 0$ in the case system follows Gaussian distribution (in which case $\left< x^4(t) \right> = 3 \left[\left< x^2(t) \right>\right]^2$). 

The models discussed so far primarily focus on one-dimensional systems. Extending these models to three-dimensional (3D) systems, as discussed in this article, is relatively straightforward. Two key considerations in this extension are: (i) the independence of diffusion processes in each direction, and (ii) the isotropic symmetry of the diffusion process in 3D, resulting in a radially symmetric process. Therefore, in the subsequent sections, we shall extend these 1D models and apply them to systems that experimentally probed in 3D systems. 

The SISF also plays a vital role in connecting theory, simulations and experiments. Using the molecular trajectories obtained in classical MD simulation, it is possible to extract the SISF directly using the formula,
\begin{equation}
    \label{sisf md}
    I_{s}(Q,t) = \frac{1}{N}\sum_{i=1}^N 
    \overline{ \left< \exp \left\{ i \mathbf{Q}. \left[\mathbf{r}_i(t+t_0) - \mathbf{r}_i(t_0)\right]  \right\} \right>_{t_0}}
\end{equation}
where $\mathbf{Q}$ is the scattering vector or momentum transfer in 3D. The overline denotes averaging over all $\mathbf{Q}$-orientations to obtain isotropic averaging. The $I_s(Q,t)$ calculated from the above formula is related to the time-Fourier transform of incoherent quasielastic neutron scattering (QENS) data \cite{Bee_1988}, $S_{inc}(Q,E)$ and hence directly provides a method to compare the simulation and experimental results. This enables QENS as a suitable technique to probe molecular self-diffusion in a variety of complex fluids \cite{Bee_1988, Arbe_1998,Arbe_2002,Burankova_2014, busch_2010, Sharma_2010, Sharma_2012, Srinivasan_2018, Srinivasan_2020}.

In this article, we employ a combination of QENS data and molecular dynamics (MD) simulations to uncover evidence of breakdown in Brownian motion behavior in molecular diffusion within complex fluids. Before delving deeper, it is crucial to briefly discuss the nature of QENS data concerning Brownian diffusion processes. As established earlier, the solution to the Brownian diffusion in 3D can be easily given as,
\begin{equation}
    \label{3d diffusion}
    P(\textbf{r},t) = \frac{1}{\sqrt{4\pi D t}} e^{-\frac{\mathbf{r}^2}{4 D t}}
\end{equation}
Upon Fourier-transformation into $(\mathbf{Q},E)$ space, we can obtain the respective signal obtained in QENS data,
\begin{equation}
\begin{split}
    \label{brownian sqe}
    S_{inc}(\mathbf{Q},E) =\frac{1}{\pi} \frac{\Gamma(Q)}{\Gamma(Q)^2 + E^2}
\end{split}
\end{equation}
where $\Gamma(Q)$ is the half-width at half-maximum (HWHM) of the Lorentzian and varies quadratically with $Q$, given by $\Gamma(Q) = DQ^2$ (in natural units, where $\hbar =1$). Hence, in systems demonstrating Brownian motion, fitting their QENS spectra with a Lorentzian and characterizing the $Q$-dependence of the HWHM of Lorentzian with quadratic variation can yield estimations of the diffusivity, $D$, of the particles within the medium.

The Lorentzian lineshape of $S_{inc}(\mathbf{Q},E)$ and the quadratic $Q$-dependence of the HWHM are the hallmark signatures of Brownian motion within the lens of QENS experiments. Therefore, any violations in either or both of these behaviours indicate a departure from the model of Brownian motion, as we shall discuss in various examples in the subsequent sections.

\section{Non-Markovian diffusion processes}

Molecular diffusion in complex fluids such as ionic liquids, molecular self-assembled aggregates, deep eutectic solvents etc., exhibit strongly heterogeneous dynamics \cite{Srinivasan_2018, Aoun_2015}. This heterogeneity is manifest in a variety of forms such as non-linear time-dependence of MSD, breakdown of Stokes-Einstein relationship, non-exponential relaxation profiles etc. In such scenarios, it is highly plausible that the underlying Markovian assumption needs to be relaxed. In order to describe a system which is inherently non-Markovian, we consider an extenstion of the Langevin equation (eq. \eqref{Langevin}), with a memory kernel, $M(t)$,
\begin{equation}
    \label{GLE}
    m\frac{dv}{dt} = -\int\displaylimits_0^t dt' M(t-t')v(t') + \zeta(t)
\end{equation}
where $\zeta(t)$ is random force which has an underlying Gaussian distribution. Eq. \eqref{GLE} is called the generalized Langevin equation (GLE). In order to respect conditions of thermodynamic equilibrium, fluctuation-dissipation relation for GLE sets a constraint between the memory kernel $M(t)$ and autocorrelation of the random force $\zeta(t)$,
\begin{equation}
    \label{FDT}
    \left< \zeta(t) \zeta(t')   \right> = k_B T M(t-t')
\end{equation}
It is easy to verify that the Langevin equation \eqref{Langevin} is recovered by considering the memory kernel of the form $M(t)=2\gamma\delta(t)$. This essentially corresponds to no memory in the random force effectively leading to the Markovian behaviour in velocity. 

A general expression for the velocity autocorrelation (VACF) for particles obeying GLE can be obtained from eq. \eqref{GLE}. Multiply on both sides of eq. \eqref{GLE} by $v(0)$ and take the ensemble average. By noting that $\left< v(0)\zeta(t)\right> = 0$ due to causality, we find that
\begin{equation}
    \label{GLE_VACF}
    \frac{d C_v(t)}{dt} = - \frac{1}{m} \int\displaylimits_0^t M(t-t') C_v(t') dt'
\end{equation}
The calculation of VACF is a convenient way to characterize the diffusion process. Further, through the Green-Kubo relation (eq. \eqref{green-kubo}) it provides a way to calculate the mean-squared displacement (MSD) as well. Considering the Laplace transform of the above equation \eqref{GLE_VACF} can serve as a mathematically convenient way to calculate the VACF for any given memory function, we can define
\begin{equation}
    \label{GLE_VACF_laplace}
    \widetilde{C}_v(u) = \frac{C_v(0)}{u + \widetilde{M}(u)/m}
\end{equation}
where $\widetilde{C}_v(u)$ and $\widetilde{M}(u)$ are the Laplace transforms of $C_v(t)$ and $M(t)$ respectively. We shall consider two particular two cases of memory kernel - exponential and inverse power-law. 

\subsection{Exponential memory kernel}
In the case of exponential memory kernel, $M(t) = \lambda^2 e^{-\lambda t}$ and it's respective Laplace transform is $\widetilde{M}(u) = \lambda^2/(u+\lambda)$. Therefore,   it is easy to obtain the VACF in the Laplace space using eq. \eqref{GLE_VACF_laplace},
\begin{equation}
    \label{VACF_exp_kernel_laplace}
    \widetilde C_v(u) = C_v(0) \frac{u+\lambda}{(u+\alpha_1)(u+\alpha_2)}
\end{equation}
where $\alpha_1$ and $\alpha_2$ are the roots to the quadratic equation ($u^2 + u\lambda+\lambda^2=0$). Inverting the Laplace transform in the above equation, we obtain
\begin{equation}
    \label{VACF_exp_kernel}
    C_v(t) = \frac{k_B T}{m(\alpha_1 - \alpha_2)} \left[ (\lambda - \alpha_1)e^{-\alpha_1 t} - (\lambda - \alpha_2) e^{-\alpha_2 t} \right]
\end{equation}
This expression clearly presents a distinct form for the VACF when compared to the result obtained through the Langevin equation (eq. \eqref{Langevin_VACF}). The two competing exponentials indicate that the VACF attains negative values which is characteristic of particles backscattering due to large number of collisions. Physically, this is a picture of transient caging of particles by it's neighbours which is captured by an exponentially decaying memory kernel in the present case. Notably, the effects of caging is the manifest feature of non-Markovianity in the system. Despite having a strikingly different behaviour from the Langevin model at short times, it can be shown that at long times the system can be described well in the Markovian approximation. This is clearly evident from MSD, which at long times ($\lambda t \gg 1 $) behaves as $\left< x^2(t) \right> \sim 2 \frac{k_B T}{\lambda m} t$, just like a particle executing Brownian motion. 
\subsection{Power-law memory kernel}
The power-law memory kernel offers a very different long-time behaviour of MSD and distinctly exhibits a non-Markovian behaviour even at very long times. In this case, $M(t) = B(\tau/t)^\beta$ where $\beta$ is power-law exponent  $\left(0<\beta<1\right)$ and $\tau$ is characteristic relaxation time. In the Laplace space, the VACF can be calculated to be,
\begin{equation}
    \label{vacf_power_law_laplace}
    \widetilde C_v(u) = C_v(0) \frac{u^{1-\beta}}{u^{2-\beta} + B\Gamma(1-\beta)\tau^\beta}
\end{equation}
where $\Gamma(z)$ is the Gamma function. The inverse Laplace transform of the above expression is linked to a special function called as the Mittag-Leffler function, $\mathrm{E}_\alpha(-t^\alpha)$,
\begin{equation}
    \label{vacf_power_law}
    C_v(t) = \frac{k_B T}{m} E_{2-\beta}\left(-B\Gamma(1-\beta)t^{2-\beta}\right)
\end{equation}
where the Mittag-Leffler functions of index $\alpha$ is defined as,
\begin{equation}
    \label{ML_func}
    E_{\alpha}(-t^\alpha) = \sum_{n=0}^\infty \frac{t^{\alpha n}}{\Gamma\left(\alpha n + 1\right)}
\end{equation}
The asymptotic large value behaviour of Mittag-Leffler functions are governed by a decaying power-law $t^{-\alpha}$. Incorporating this in eq. \eqref{vacf_power_law}, we obtain the long-time behaviour of the VACF in the case of a power-law memory kernel to be,
\begin{equation}
    \label{vacf_power_law_lt}
    C_v(t) \xrightarrow{t\rightarrow\infty}\frac{sin(\pi\beta)}{\pi\beta}\beta(\beta-1) \left(\frac{t}{\tau}\right)^{\beta-2}
\end{equation}
The above expression suggests that VACF is necessarily negative and decays to zero as a power-law, which is typically much slower decay than an exponential decay observed in the case of Langevin equation. This slow power-law decay even at long times makes sure that the system is distinctly non-Markovian always. This is also reflected in the behaviour of MSD, which in the long-time limit is not linear in time, but sublinear, $\left< x^2(t) \right> \sim t^\beta$. This behaviour is also called subdiffusion, being strongly slower than the rate at which Brownian particle diffuses. The power-law memory kernel holds a special importance in the study of non-Markovian processes, as it is linked to definition of fractional Brownian motion (fBm), which will be discussed in detail in the penultimate section of this article. 

\subsection{Lateral subdiffusion of lipids}

The lateral motion of the lipid within the leaflet is of key interest since it plays an important role in various physiologically relevant membrane processes including cell signalling, membrane trafficking, and cell recognition. However, the model description of the lateral motion of the lipids is not fully agreed upon in the literature. Different models such as Fickian diffusion \cite{Armstrong_2011, Sharma_2015, Sharma_2016, Dubey_2018, Dueby_2019}, ballistic flow like motion \cite{busch_2010, falck_2008}, localized translational motion of lipids in a cylindrical volume \cite{wanderlingh_2014}, and sub-diffusive motions \cite{Flenner_2009,Jeon_2012, Srinivasan_2018, Samapika_2023}, have been used to describe the lateral motion of the lipid. In this section, we show how the lateral motion of lipids can be described within the framework of generalized Langevin equation (GLE) with a power-law memory kernel, for a lipid bilayer system made of cationic lipids dioctadecyl dimethyl ammonium bromide (DODAB)\cite{Srinivasan_2018, Sharma_2020, Samapika_2023}. The results on DODAB bilayer is presented at two different temperatures, 298 K and 350 K, corresponding to ordered and the fluid phases of the DODAB membrane system respectively.

\subsubsection{Results from MD simulations}
\begin{figure}
    \includegraphics[width=0.65\linewidth]{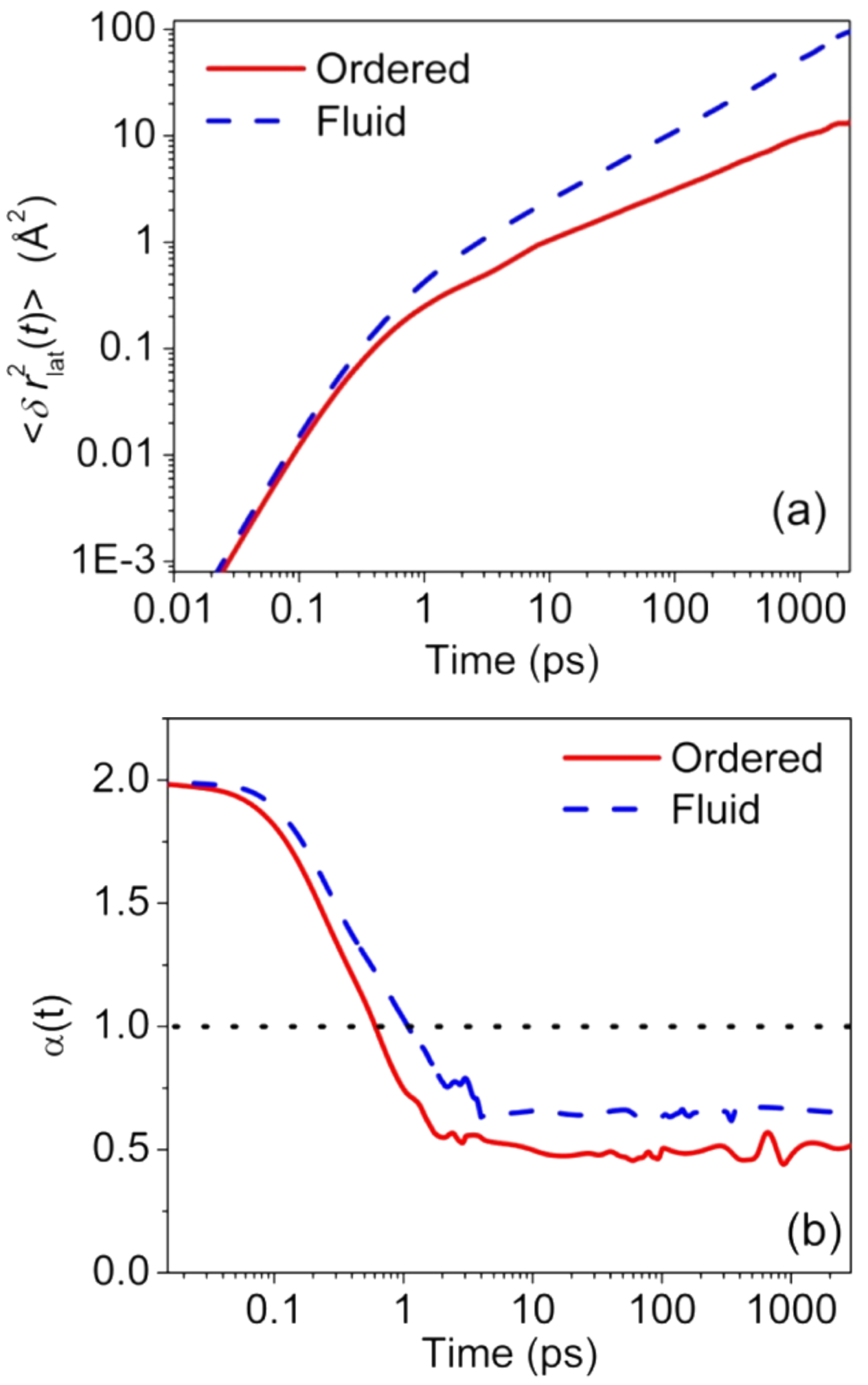}
    \caption{(a) Lateral MSD of DODAB lipids in the ordered and fluid phases. (b) The corresponding subdiffusive exponents for the lipids in both the phases. Figure adapted from ref. \cite{Srinivasan_2018} }
    \label{msd dodab}
\end{figure}

Fig. \ref{msd dodab}(a) shows the plot of lateral MSD for both ordered and fluid phases upto $\sim$ 2 ns as calculated from MD simulations \cite{Srinivasan_2018} on dioctadecyldimethyl ammonium bromide (DODAB) bilayer at two different temperatures of 298 K and 350 K corresponding to ordered and fluid phases. It is observed that, at short times, both phases show ballistic behavior with $t^2$ dependence, which is typically timescale before any interaction with neighbouring lipids. This regime is followed by the subdiffusive regime – described by a power law dependence of $t^\alpha \; (\alpha < 1)$. The explicit dependence of power law can be calculated by the following formula,
\begin{equation}
    \label{msd power}
    \left< \delta r^2_{lat}(t) \right> = At^\alpha \quad \implies \quad \alpha(t) = \frac{d \left[\ln \left< \delta r^2_{lat}(t) \right>\right]}{d \left[\ln t\right]}
\end{equation}
Fig. \ref{msd dodab}(b) shows the variation of $\alpha$ with respect to time t, indicating sub-diffusion in both the ordered ($\alpha \sim$ 0.5) and fluid ($\alpha \sim$ 0.62) phases with $\alpha$ < 1, wherein a value of $\alpha$ = 1 indicates Brownian diffusion. Therefore, quite clearly the lateral motion of lipids is strongly subdiffusive in both the ordered and fluid phases. Notably, a stronger breakdown of Brownian motion is observed in the ordered phase, which is likely linked to the denser packing of lipids. This suggests that the sub-diffusive motion of lipids in the bilayer can be associated to crowding of lipids in the system, which can be modeled as a non-Markovian diffusion process \cite{Metzler_2016, Jeon_2012}.

\begin{figure}
    \centering
    \includegraphics[width=0.65\linewidth]{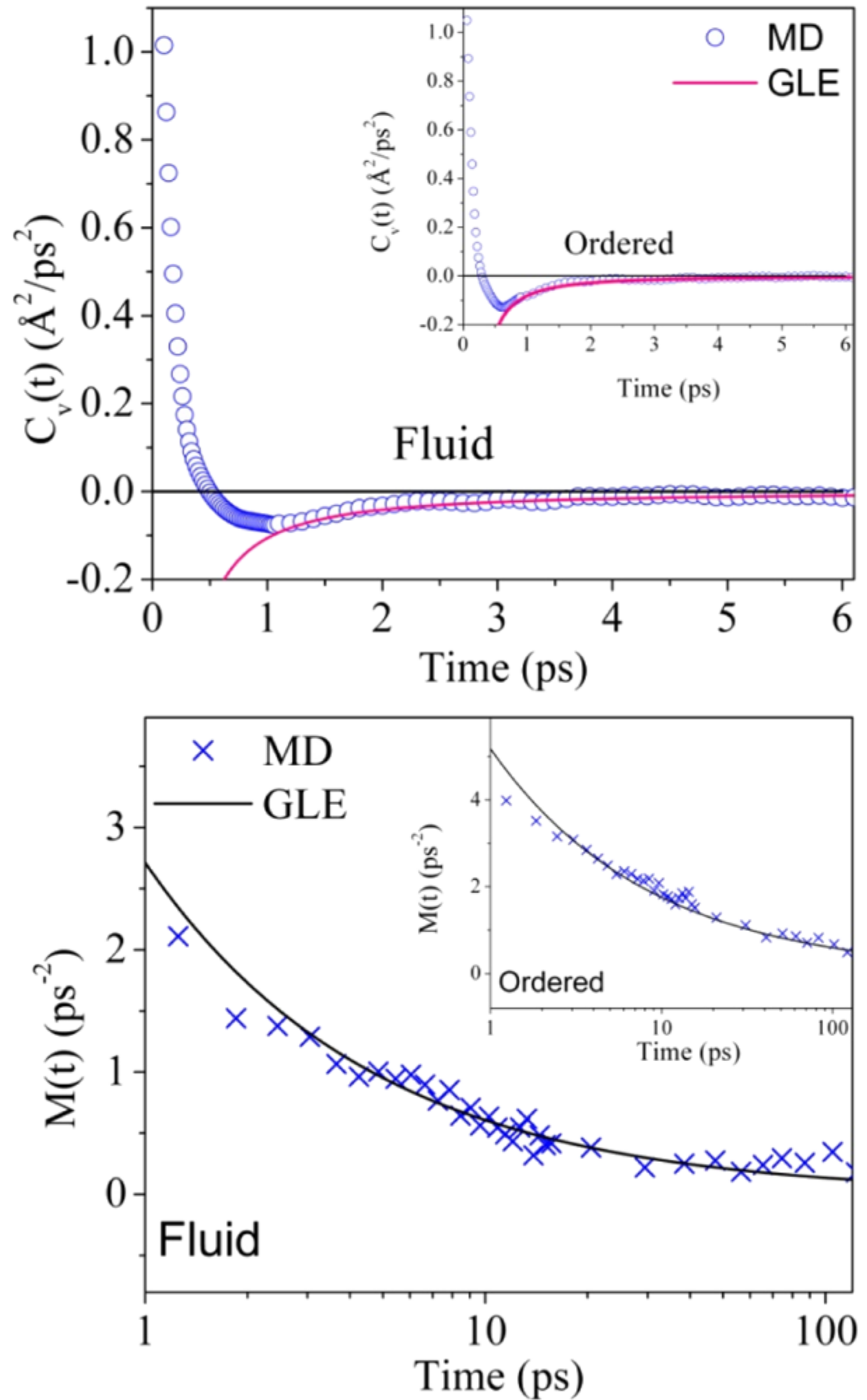}
    \caption{(a) The VACF of DODAB lipids calculated from MD simulation trajectories in the fluid and ordered (inset) phases. (b) The corresponding memory functions calculated from eq. \eqref{GLE_VACF} using VACF in each phase. The solid lines indicate the theoretical fits based on eq. \eqref{asymptotic limits}.   Figure adapted from ref. \cite{Srinivasan_2018} }
    \label{vacf mem}
\end{figure}

The framework of the generalized Langevin equation (GLE) with Gaussian colored noise, as described in \cite{gardiner2009}, is well-suited for modeling the lateral subdiffusion of lipids. In Fig. \ref{vacf mem} (a), the simulated velocity autocorrelation function ($C_v(t)$) for fluid and ordered phases (inset) is depicted with open symbols. As detailed in Section 2, $C_v(t)$ is connected to the memory function ($M(t)$) through an integro-differential equation given by \eqref{GLE_VACF}. This equation is utilized to numerically compute $M(t)$ from the simulated $C_v(t)$ of lipid center of mass (COM). In Fig. \ref{vacf mem} (b), the computed behavior of the memory function $M(t)$ over time is shown for both fluid and gel phases. It is observed that both $C_v(t)$ and $M(t)$ decay more rapidly in the fluid phase compared to the ordered phase due to the higher dynamics present in the former. This methodology allows for a detailed analysis of lipid dynamics, specifically capturing the subdiffusive behavior, and provides insights into phase-dependent variations in the memory function and VACF. 

The subdiffusion behaviour of the lipid COM can be described by considering a power-law memory kernel which essentially leads to a fractional Brownian motion in the long-time limit. In this scenario, where the MSD of the lateral motion obeys eq. \eqref{msd dodab}, the theoretical asymptotic behavior for the VACF is given by eq. \eqref{vacf_power_law_lt} and the associated memory function in the asymptotic limit is given by,

\begin{equation}
    \label{asymptotic limits}
    M(t) \xrightarrow{t\rightarrow\infty} \frac{\left<v^2\right>}{A} \frac{\sin (\pi \alpha)}{\pi \alpha} t^{-\alpha}
\end{equation}
The solid lines in Fig. \ref{vacf mem}(a) and \ref{vacf mem}(b) indicate the theoretical functions obtained for $C_v(t)$ and $M(t)$ in the asymptotic limit ($t > 1$ ps), with values of $\alpha$ fixed at 0.5 and 0.62 for ordered and fluid phases respectively. The values of A are found to be 0.36 $\text{\text{\AA}}^2\text{/ps}^\alpha$( $\alpha$ = 0.5) and 0.41 $\text{\text{\AA}}^2\text{/ps}^\alpha$ ($\alpha$ = 0.62) in the ordered and fluid phases, respectively. The excellent quality of the fits indicates that GLE with a power-law memory kernel is a good description of the lateral diffusion of lipids. Based on eq. \eqref{asymptotic limits}, the memory function decays faster for a larger value of $\alpha$, hence suggesting the memory effects are longer lived in the ordered phase compared to the fluid phase. 

\subsubsection{Lateral subdiffusion - validation through QENS}

\begin{figure}
    \centering
    \includegraphics[width=0.65\linewidth]{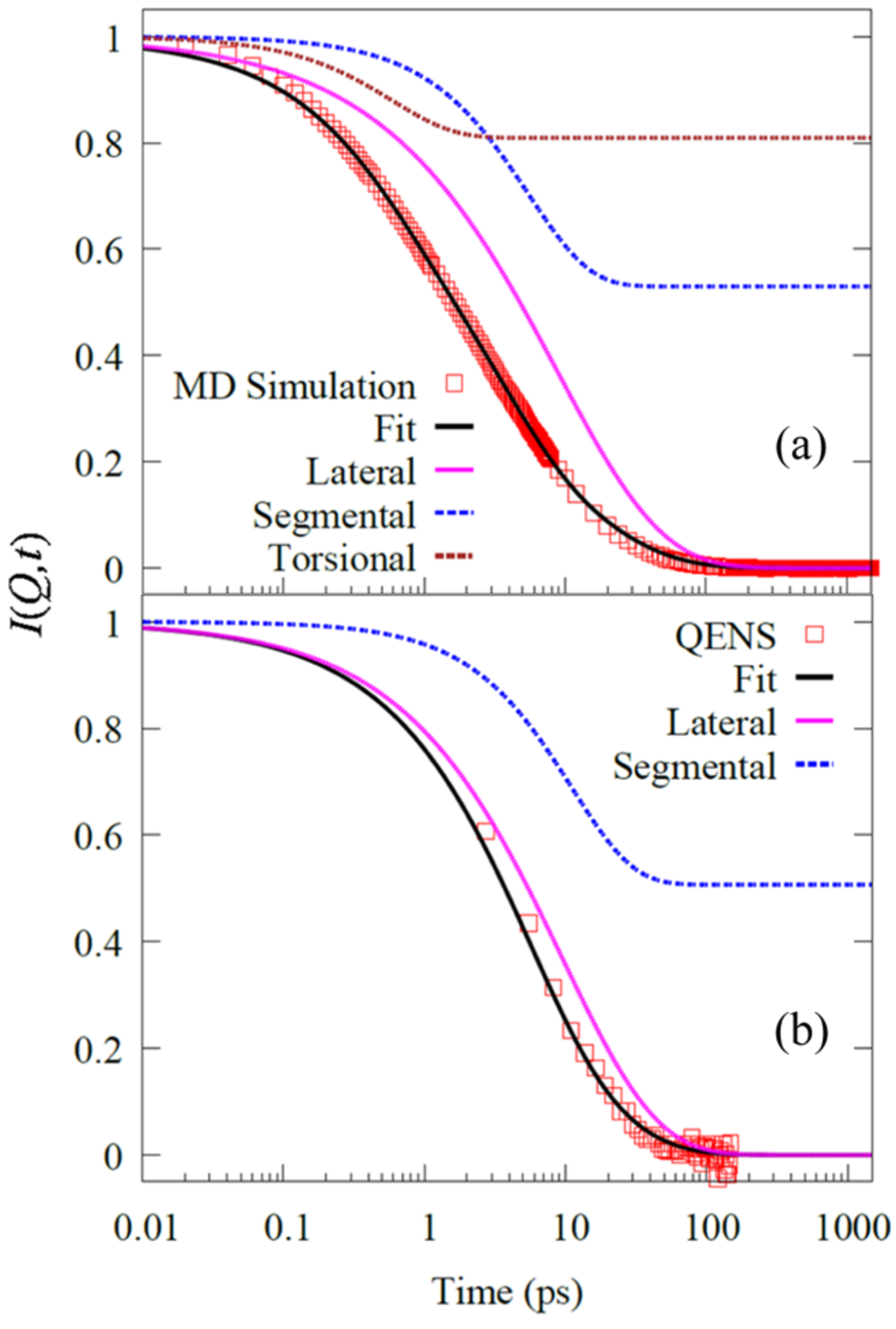}
    \caption{All hydrogen SISF, $I(Q,t)$, from (a) MD simulation trajectories (b) Fourier transform of QENS spectra at $Q$=1.2 $\text{\AA}^{-1}$. The fits based on eq. \eqref{iqtH md} and eq. \eqref{iqtH qens} for MD and QENS are respectively indicated along with their respective components.  Figure adapted from ref. \cite{Srinivasan_2018}}
    \label{iqt dodab}
\end{figure}

The sub-diffusive nature of lateral motion in DODAB lipids, as observed in molecular dynamics (MD) simulations, can be further confirmed experimentally through quasielastic neutron scattering (QENS) data analysis. By modeling the dynamics of the DODAB lipid bilayer in the fluid phase, comparisons can be drawn between MD simulation results and experimental QENS data. The incoherent neutron scattering law is linked to the self-intermediate scattering function (SISF) of the system, predominantly involving hydrogen atoms, through a time-Fourier transform. MD simulation trajectories can be used to calculate the all hydrogen SISF, $I_H(Q,t)$. The motion of hydrogen atoms is a combination of lateral and internal motions. The QENS data is modelled assuming lateral and segmental dynamics of lipids, however MD simulations indicates an extra dynamical component, faster torsional motions. Considering these three degrees of freedom, the model for fitting the MD simulation SISF can be given as,
\begin{equation}
    \label{iqtH md}
    I_H(Q,t) = e^{\left(-\Gamma_{lat} t^\alpha\right)} \left[ a_0 + (1-a_0) e^{-\Gamma_{seg} t } \right] \left[ b_0 + (1-b_0) e^{-\Gamma_{tor} t } \right]
\end{equation}
where the first term corresponds to the lateral motion, considering spatially homogenous subdiffusion as described in the framework of GLE. The second and third terms correspond to segmental and torsional motions. The details of all the components are described here \cite{Srinivasan_2018}. In this section we focus, primarily only on the lateral component of the dynamics. The characteristic relaxation time of torsional motions ($\Gamma_{tor}$) is ~ 1 ps, making it too fast to be observed within the energy transfer range of the IRIS spectrometer. Therefore, the corresponding model for fitting the SISF obtained from the time-Fourier transform of QENS data can be written as,
\begin{equation}
    \label{iqtH qens}
    I_{QENS}(Q,t) = e^{\left(-\Gamma_{lat} t^\alpha\right)} \left[ a_0 + (1-a_0) e^{ (-\Gamma_{seg} t } \right]
\end{equation}
The all-hydrogen SISF calculated from MD simulation (350 K) and QENS data (345 K) at a representative $Q$=1.2 $\text{\AA}^{-1}$ are shown in Fig. \ref{iqt dodab}. The model fits based on eq. \eqref{iqtH md} and eq. \eqref{iqtH qens} for MD and QENS respectively are also shown in the plots, along with the individual components. The quality of the fits indicates that this model successfully describes the dynamics of the membrane in fluid phase. 
 
\begin{figure}
    \centering
    \includegraphics[width=0.65\linewidth]{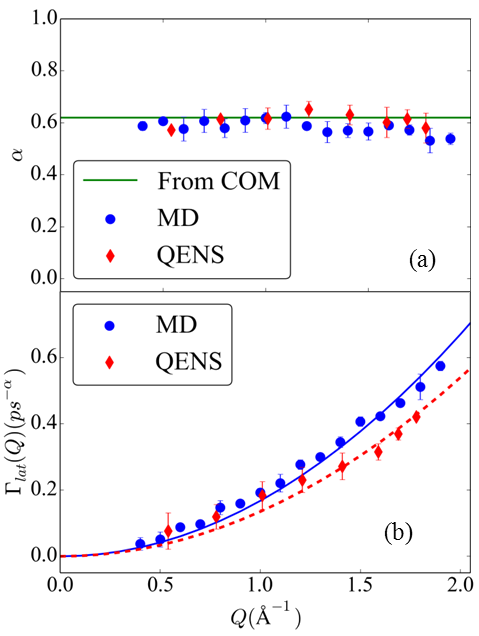}
    \caption{Variation of the (a) the exponent, $\alpha$, and (b) $\Gamma_{lat}$ with respect $Q$ as obtained from fitting to SISF obtained from MD simulation and QENS data. Solid line in (a) indicates the value obtained from the MSD of lipid COM motion (eq. \eqref{msd dodab}). The continuous (MD) and dashed (QENS) lines in (b) correspond to the respective quadratic fits using $(A/4)Q^2$.  Figure adapted from ref. \cite{Srinivasan_2018}}
    \label{gamma lateral}
\end{figure}

The lateral motion of the lipids is characterized by the exponent of sub-diffusion ($\alpha$) and the associated relaxation time ($1/\Gamma_{lat}$). The varation of $\alpha$ from the fits of QENS and MD data are shown in Fig. \ref{gamma lateral}(a). The $Q$-averaged value of the sub-diffusive exponent, $\alpha$, from both the simulation and experiment is $\sim$ 0.61, which is very close to the exponent obtained in the GLE description for lipid COM from MD simulations. The variation of the relaxation timescale associated to lateral motion, $\Gamma_{lat}$ is shown Fig. \ref{gamma lateral}(b) for both QENS and MD simulation. The solid lines indicate fitting based on quadratic dependence $\frac{1}{4} AQ^2$, considering a Gaussian diffusion process – as observed from the analysis of lipid COM trajectories. The value of $A$ obtained from the least-squares fit was found to be 0.42 $\text{\text{\AA}}^2\text{/ps}^\alpha$( $\alpha$ = 0.61) and 0.34 $\text{\text{\AA}}^2\text{/ps}^\alpha$( $\alpha$ = 0.61) for MD simulation and QENS experiments respectively. The obtained values are comparable to their counterpart obtained in the GLE description of lipid COM.

This analysis establishes the non-Markovian nature of the lateral motion of lipids in a bilayer membrane. Further, it also shows that the diffusion is truly Gaussian and can be described within the framework of generalized Langevin equation (GLE) using a power-law memory kernel. It should be noted that this model is also equivalent to fractional Brownian motion (fBm) in the asymptotic long time limit, which shall discussed in that penultimate section.

\section{Non-Gaussian diffusion processes - Jump diffusion}
While GLE provides a robust framework to describe inherent non-Markovian nature of diffusion process, it is still fundamentally built on Gaussian noises. Therefore, these models cannot describe non-Gaussian diffusion processes. In order to mathematically describe such processes, we propose a non-local diffusion (NLD) equation given by,
\begin{equation}
    \label{non-local diffusion}
    \frac{\partial P(x,t)}{\partial t} = \int\displaylimits_{-\infty}^{\infty} dx' \; \Lambda_h(x-x') \frac{\partial^2 P(x',t)}{\partial x'^2}
\end{equation}
where $\Lambda_h(x)$ is the jump kernel that incorporates non-local displacements into the original diffusion equation. Considering only infinitesimally small local displacements based on a Dirac delta jump-kernel, i.e. $\Lambda_h(x) = D \delta(x)$, leads to the case of Brownian motion described by the original diffusion equation (eq. \eqref{diffusion}). This implies that if the condition of non-local displacements is relaxed we obtain the original diffusion equation. Meanwhile, choosing a power-law kernel, $\Lambda_h(x) \sim x^{-\alpha}$ leads to the description of Levy flights, which is strongly non-Gaussian at all length scales. In our study we aim to develop jump kernels which exhibit non-Gaussianity over small length scales but tend to become Gaussian at large spatial distances, reverting to the hydrodynamic regime.

Before the development of the general jump-kernel, we note that the general solution to eq. \eqref{non-local diffusion} is given easily in Fourier space, in terms of SISF, $I(k,t)$. Considering the Fourier transform of eq. \eqref{non-local diffusion} leads to an ordinary differential equation in time whose solution is given by,
\begin{equation}
    \label{non-local diffusion FT}
    I(k,t) = I(k,0) \exp\left[-k^2 \hat\Lambda_h(k) t \right]
\end{equation}
where $\hat \Lambda_h(k)$ is the Fourier transform of the jump kernel $\Lambda_h(x)$. Therefore, the solutions of the NLD equation can in principle be obtained by calculating the inverse Fourier transform of eq. \eqref{non-local diffusion FT}. 

\subsection{Generalized jump-kernel}
The general form of jump-kernel we require in our study should exhibit transient non-Gaussian effects at short distances, and revert to Gaussian behaviour at long distances. This sets some strong conditions to be imposed on the jump-kernel. In the Fourier domain, these physical conditions can be simplified into two specific limiting rules as indicated below,
\begin{equation}
    \label{limit_jump_rules}
    \begin{split}
        \hat\Lambda_h(k) \xrightarrow{k\rightarrow 0} \frac{x_0^2}{\tau_j} \propto D_j \qquad \qquad
        \hat\Lambda_h(k) \xrightarrow{k\rightarrow\infty} \frac{k^{-2}}{\tau_j}
    \end{split}
\end{equation}
Here $x_0$ and $\tau_j$ are related to characteristic \textit{jump length} and \textit{mean waiting time between jumps} of the jump diffusion process, respectively. The ratio between these two parameters is also related to the jump diffusivity, $D_j$. The first condition ($k\rightarrow 0$) pertains to the long wavelength limit, which is associated to the diffusion behaviour at long distances, wherein the system reverts to Gaussian diffusion behaviour with SISF following $I(k,t) = e^{-D_j k^2 t}$. In order to make sense of the second condition ($k\rightarrow\infty$), let us consider a typical SISF of jump diffusion process of the form $I(k,t) = e^{-t/\tau(k)}$, where $\tau(k)$ is the wavevector-dependent relaxation time. In the short-wavelength limit ($k\rightarrow \infty$), the behaviour of $\tau(k)$ for jump diffusion process saturates to a constant value of the average waiting time between jumps, $\tau_j$. Therefore, this demands that in this limit, $\hat\Lambda_h(k)$ asymptoticaclly decays as $k^{-2}$, which is given as the second condition in eq. \eqref{limit_jump_rules}. In addition to the conditions explicitly given in eq. \eqref{limit_jump_rules}, the isotropic symmetry dictated by the diffusion problem also ensures that $\Lambda_h(k)$ is an even function of $k$.

Having these conditions (eq. \eqref{limit_jump_rules}) explicitly , we propose a generalized jump-kernel using a series expansion with respect to $kx_0$,
\begin{equation}
    \label{general jump-kernel}
    \Lambda_h(k) = \frac{k^{-2}}{\tau_j} \sum_{n=1}^{\infty} c_n (kx_0)^{2n}
\end{equation}
where the choice of coefficients $c_n$ is dictated by satisfying the conditions in eq. \eqref{limit_jump_rules}. Considering only the even powers in expansion is a consequence of isotropic symmetric in the diffusion process. The values of $\{c_n\}$ is such that the summation converges to unity when $kx_0 \rightarrow \infty$ and the sum is proportional to $x_0^2/\tau_j$ for $kx_0 \rightarrow 0$. A typical example is choosing $c_n = (-1)^{n-1}$, which leads to a jump-kernel of the form $\hat \Lambda_h(k) = (x_0^2/\tau_j)\left[1+\left(kx_0\right)^2\right]^{-1}$ in the Fourier space. Notably, this is equivalent to considering a symmetric exponential jump-kernel, i.e., $\Lambda_h(x) = \left(2\tau_j x_0\right)^{-1} e^{-|x|/x_0}$. Different possible choices of $\{c_n\}$ and the related jump-kernels are listed in Table 1.1, along with their analytic inverse Fourier transform in some cases. It is crucial to note that the jump diffusivity, $D_j = c_1 (x_0^2/\tau_j) $ and therefore depends on the definition of the jump kernel. Employing the generalized jump-kernel in eq. \eqref{non-local diffusion FT}, we have a general form for SISF of jump diffusion process,
\begin{equation}
    \label{general jump SISF}
    I(k,t) = \exp \left[ - \frac{t}{\tau_j} \sum_{n=1}^\infty c_n (kx_0)^{2n} \right]
\end{equation}
where we have set $I(k,0)=1$ which is tantamount to the initial condition $P(x,0) = \delta(x)$.

\bgroup
\def\arraystretch{2.5}
\begin{table*}
\label{jump kernel list}
    \centering
    \caption{List of various jump-kernels from the generalized formula \eqref{general jump-kernel}. *$E_{2n}$ is the sequence of Euler numbers}
    \begin{tabular*}{\textwidth}{@{\extracolsep{\fill}}lll}
        \hline
        $\{c_n\}$ &
        $\hat\Lambda_h(k)$  & 
        \(\displaystyle \Lambda_h(x) = \frac{1}{2\pi} \int_{-\infty}^{\infty} e^{-ixk} \hat\Lambda_h(k) dk \) \\
        \hline\hline
         $c_n = \left(-1\right)^{n-1}$ & 
         \(\displaystyle \frac{1}{\tau_j} \frac{x_0^2}{1+(kx_0)^2}\)  &
         \(\displaystyle \frac{1}{2\tau_j x_0} \exp\left[-|x|/x_0\right] \) \\
         
         \(\displaystyle c_n = \frac{\left(-1\right)^{n-1}}{n!} \) & 
         \(\displaystyle \frac{1}{\tau_j} \frac{1-e^{-(kx_0)^2}}{k^2} \) &
         \(\displaystyle \frac{1}{2\tau_j} \bigg(x \mathrm{Erf}\left[x/(2x_0)\right] - |x| \bigg) + \frac{x_0^2}{\tau_j\sqrt{\pi}} e^{-x^2/(4x_0^2)} \) \\
         
         \(\displaystyle c_n = \frac{(-1)^{n-1}}{(2n+1)!}   \) &
         \(\displaystyle \frac{1}{\tau_j} \frac{ 1 - \frac{\sin (kx_0)}{kx_0} }{k^2} \) &
         \(\displaystyle \frac{\left(x_0 -x\right)|x_0 - x| + (x_0 + x)|x_0 + x|}{8\tau_j x_0} - \frac{|x|}{2\tau_j} \)\\
         
         \(\displaystyle c_n = \frac{(-1)^{n-1}}{n+1} \) &
         \(\displaystyle  \frac{k^{-2}}{\tau_j} \left[1 - \frac{\ln\left(1+(kx_0)^2\right)}{(kx_0)^2} \right] \) &
         No analytic form\\
         
         \(\displaystyle c_n = \frac{E_{2n}}{(2n)!}  \) *& 
         \( \displaystyle \frac{k^{-2}}{\tau_j} \left[ 1 - \mathrm{sech}(kx_0) \right] \)& 
         No analytic form\\
         \hline
    \end{tabular*}
\end{table*}
\egroup

Using the eq. \eqref{moments}, we can directly compute the non-Gaussian parameter for the general SISF of jump diffusion given in eq. \eqref{general jump SISF} to be,
\begin{equation}
    \label{NG jump}
    \alpha_2(t) = \frac{-2 c_2}{c_1^2} \frac{\tau_j}{t}
\end{equation}
This expression indicates two crucial aspects - firstly non-Gaussian parameter exhibits a slow $t^{-1}$ decay with time, additionally the second coefficient is necessarily negative ($c_2 < 0$) for systems exhibiting heavier than Gaussian tails in the jump diffusion process, so that $\alpha_2(t) > 0$.

\subsection{Equivalence with Continuous Time Random Walk}
It is possible to establish a direct correspondence between the framework of continuous time random walk (CTRW) and the NLD model discussed here. To address this connection, let us consider a CTRW model wherein a particle undergoes a series of jumps separated by discrete time-intervals \cite{Montroll_1965}. The two fundamental defining elements of a CTRW model are the jump-length distribution, $\rho(x)$ and the waiting time distribution $w(\tau)$. These distributions carry information about the probabilistic nature of jumps and waiting times between jumps. While, in general the jump-lengths and waiting times may be correlated, in the present scenario we conisder the uncorrelated case which presents itself as a Markovian renewal process. For a particle executing CTRW, the PDF associated to the displacement of the particle is given by,
\begin{equation}
    \label{prob ctrw}
    P_{c}(x,t) = \sum_{n=0}^\infty \mathcal{P}(n,t)p_n(x)   
\end{equation}
where $\mathcal{P}(n,t)$ is the probability of $n$ jumps occuring before time $t$ and $p_n(x)$ is the probability that the $n$ jumps add to a distance of $x$. For the case, where the particle is executing independent jumps, it can be shown that $p_n(x) = \left[ \rho(x) \right]^{\otimes n}$, where $\otimes n$ refers to $n$ convolutions. Meanwhile, considering an exponential waiting-time distribution of the form, $w(\tau) = \tau_a^{-1} e^{-\tau/\tau_a}$  essentially leads to $\mathcal{P}(n,t)$ being a Poisson process. Therefore, we can rewrite eq. \eqref{prob ctrw}
\begin{equation}
    \label{prob ctrw 2}
    P_c(x,t) = \sum_{n=0}^\infty \frac{(t/\tau_a)^n}{n!}e^{-t/\tau_a} \left[ \rho(x)\right]^{\otimes n}
\end{equation}
Considering a space Fourier Transform of the above equation will give the SISF associated to the respective CTRW process of the form,
\begin{equation}
    \label{iqt ctrw}
    I_c(k,t) = e^{-t/\tau_a} \sum_{n=0}^\infty \frac{(t/\tau_a)^n}{n!} \left[ \hat \rho(k) \right]^n = \exp\left( -\frac{t}{\tau_a}\left[1-\hat\rho(k)\right]\right)
\end{equation}
where we have written $\hat\rho(k)$ as the Fourier transform of the $\rho(x)$. Having obtained the SISF for the CTRW under the Markovian approximation, we can compare this with the solutions obtained using the NLD equation given in eq. \eqref{non-local diffusion FT}. It clearly reveals that the a relationship between the jump-kernel $\Lambda_h$ and the jump-length distribution $\rho$ in the Fourier space,
\begin{equation}
    \label{kernel dist}
    \hat \Lambda_h(k) = \left(\frac{1}{\tau_j}\right)\frac{1 - \hat \rho(k)}{k^2}
\end{equation}
This relationship clearly shows that there is a one-one correspondence between the non-local jump diffusion equation in eq. \eqref{non-local diffusion} and Markovian CTRW model with exponential waiting time distribution. In fact, it is possible to calculate the different possible jump-length distributions directly from list of jump-kernels listed in Table 1.1. However, as we shall see in the next section, the real benefit of NLD equation model is its applicability to non-Markovian models, providing a framework diffusion in media involving both non-Markovian and non-Gaussian effects. 

\subsection{Jump diffusion in deep eutectic solvents (DESs)}
The early years of this century witnessed the emergence of a new class of solvents, known as deep eutectic solvents (DESs) \cite{Abbott_2003, Abbott_2004, Smith_2014, zhang_2012}, which exhibited physicochemical properties very similar to ILs. DESs are created by mixing two or more compounds at a specific molar ratio corresponding to their eutectic point \cite{Abbott_2003, Abbott_2004, Smith_2014, zhang_2012, Guchhait_2014, Boisset_2013}. Generally, these mixtures have a considerably lower freezing point compared to the parent compounds \cite{Smith_2014, zhang_2012}. These solvents have found extensive applications in various industrial processes, including electrodeposition \cite{Lin_1999, Smith*_2013, abbott_2012, Sun_2012, Gomez_2011}, catalysis \cite{Chirea_2011}, nanoparticle \cite{Chirea_2011}, and nanotube \cite{Liu_2010, Gutierrez_2011} synthesis, drug transport \cite{Morrison_2009}, $\text{CO}_2$ capture \cite{Xie_2016}, etc.

Recently, it has been shown that a mixture of acetamide ($\text{CH}_3 \text{CONH}_2$) and a group of lithium salts (LiX, X = $\text{ClO}_4$, Br, $\text{NO}_3$) in a molar ratio of 78:22 form DES having freezing points below room temperature\cite{Guchhait_2014, Guchhait_2012}. It has been suggested that depression of freezing points in these systems can be ascribed to the ability of these salts to break the inter-amide hydrogen bonding. Among the three salts, $\text{LiClO}_4$ forms the least viscous DES at room temperature, indicating larger affinity of perchlorate ion to amide group\cite{Guchhait_2014}. While there are plenty of studies involving the macroscopic transport of DES systems, a comprehensive model of microscopic dynamics is not well established. The diffusion mechanism of acetamide in these DESs \cite{Srinivasan_2020, Srinivasan_2020a, Srinivasan_2023}is unraveled through MD simulations, and these results are utilized to model the QENS experimental data. The observed behaviour establishes the existence of caging and jumps in the diffusion process, which leads to the manifestation of dynamical heterogeneity (DH)  and non-Gaussian diffusion process.

\subsubsection{Cage-jump mechanism - MD simulation}
\begin{figure}
    \centering
    \includegraphics[width=0.65\linewidth]{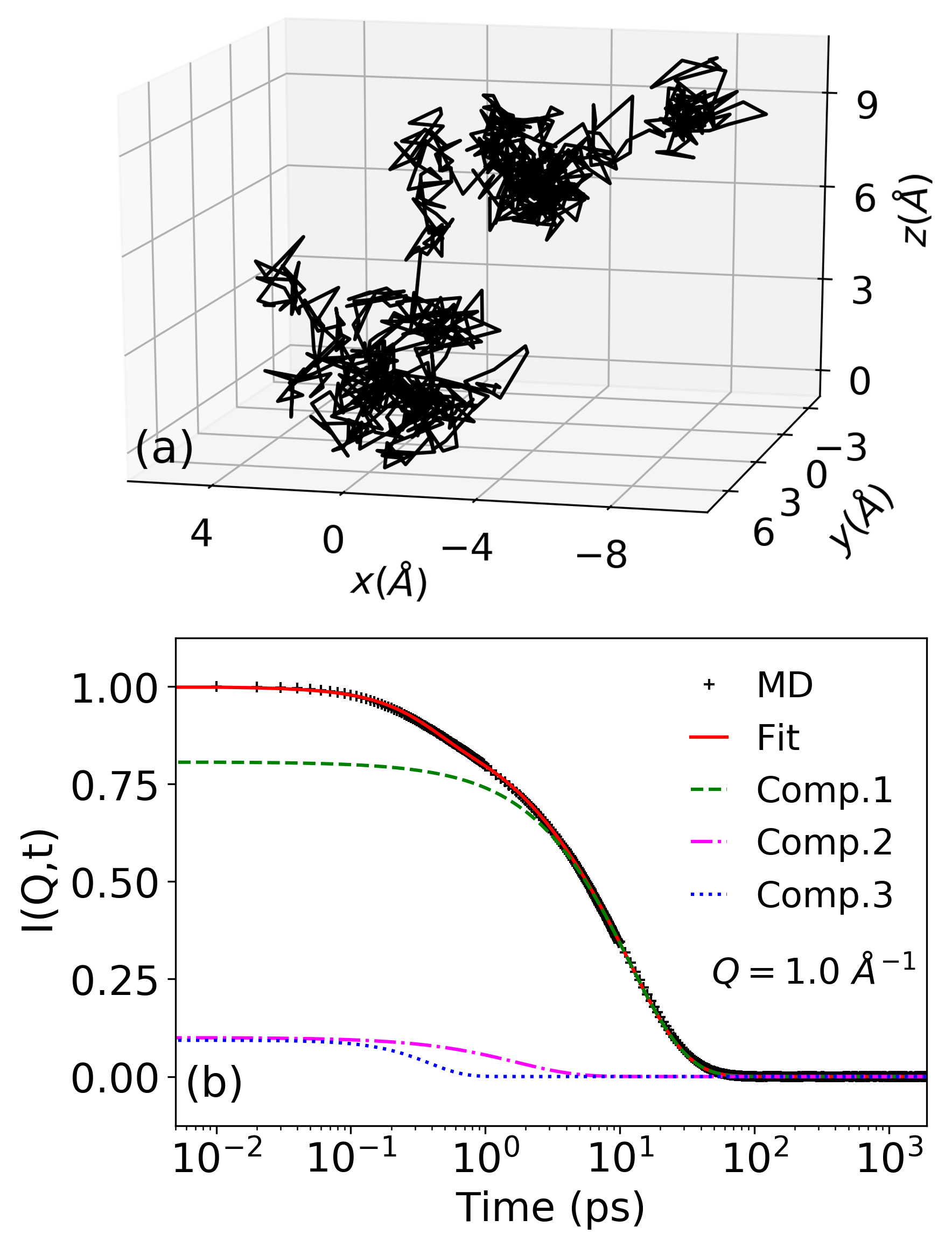}
    \caption{(a) MD simulation trajectory of acetamide molecules within the DES. The trajectory shows evident signatures of cages and jumping. (b) $I(Q,t)$ calculated from the MD simulation trajectory of acetamide COM within DES. The fits based on eq. \eqref{iisf acetamide final} along with their components are shown in the figure. The comp. 1 and comp. 2 correspond to the jump and localized diffusion components respectively, and comp. 3 corresponds to the ballistic motion.  Figure adapted from ref. \cite{Srinivasan_2020} }
    \label{traj iqt des}
\end{figure}

The MD simulation trajectory of acetamide molecules within the acetamide+ $\text{LiClO}_4$ DES is shown in Fig. \ref{traj iqt des}(a) clearly vindicating the formation of transient cages and diffusive jumps between these cages. The diffusion model describing acetamide dynamics is chosen considering three different processes – ballistic motion, localized motion inside a transient cage and cage-cage free diffusion. This is motivated by observing the trajectories in Fig. \ref{traj iqt des} (b) in which local clustering and jump-like motion are quite evident. A similar model has been employed in studying the dynamics of supercooled water by Qvist et. al. \cite{Qvist_2011}. Similar models have also been employed in describing difffusion in ionic liquids \cite{Burankova_2014, Embs_2013, Berrod_2017} and glyceline DES \cite{Wagle_2015}. Therefore the SISF corresponding to acetamide dynamics can be written as,
\begin{equation}
    \label{iisf acetamide final}
    \begin{split}
    I (Q,t) = \alpha (Q)\left[ {C_0 (Q)e^{ - \zeta _1 t}  + \left( {1 - C_0 (Q)} \right)e^{ - \left( {\zeta _1  + \zeta _2 } \right)t} } \right] \\
    + \left( {1 - \alpha (Q)} \right)e^{ - \left( {\sigma _v Qt} \right)^2 /2} 
    \end{split}
\end{equation}
where the first term in the square brackets correspond to the diffusive components of motion which include diffusion inside the transient cage and cage-to-cage free diffusion. The last term described by a Gaussian term represents the ballistic motion of the molecule. The least-squares fits of eq. \eqref{iisf acetamide final} to the calculated $I(Q,t)$ of acetamide are shown in Fig. \ref{traj iqt des} (b) $Q = 1.0 \text{\AA}^{-1}$. The individual components are also indicated in these figures. The diffusive components are indicated as component 1 \& 2 and component 3 represents the ballistic motion. The quality of fits is found to be excellent which confirms the validity of the model. It is important to note that the parameters $\zeta_1(Q)$ and $\zeta_2(Q)$ correspond to the inverse relaxation timescales associated to jump and localized diffusion processes, respectively.

\subsubsection{QENS experimental results}
\begin{figure}
    \centering
    \includegraphics[width=0.65\linewidth]{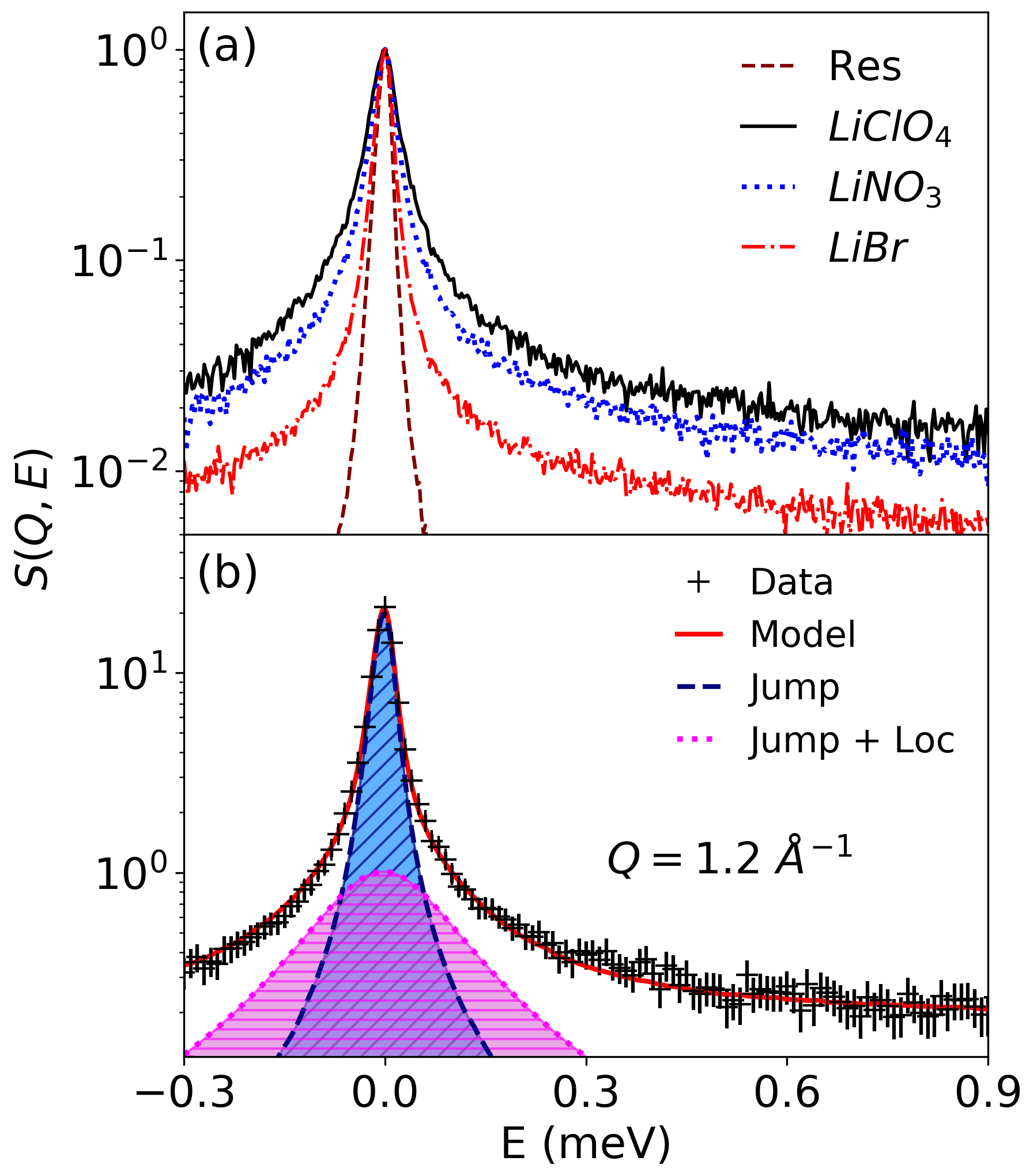}
    \caption{(a) QENS data of all three DESs at 330 K shown at $Q = 1.2 \text{\AA}^{-1}$. The data are peak-normalized to visualize the difference in QENS broadening in different systems. (b) QENS data of DES based on acetamide + $\text{LiNO}_3$ at $Q = 1.2 \text{\AA}^{-1}$, shown along with model fitting based on sum of two Lorentzians. The Lorentzians associated to jump and jump+localized diffusion processes are indicated by different shaded regions. Figure adapted from ref. \cite{Srinivasan_2020} }
    \label{qens des}
\end{figure}

Figure \ref{qens des}(a) shows the QENS data measured on three different DESs at temperature of 330 K, at a $Q$-value of 1.2 $\text{\AA}^{-1}$. It is important to note that the QENS data in all these three systems only captures the diffusion of acetamide molecules, owing to high incorhernt scattering cross-section of hydrogen atoms. From the trend of the QENS spectra, it can be evidently concluded that the diffusion of acetamide is fastest in the $\text{LiClO}_4$ DES and slowest in in LiBr DES. The QENS data is described based on the cage-jump diffusion model, which has been established through a comprehensive investigation from the MD simulation study.  Notably, the ballistic component of the motion is too fast to be detected within the experimental window of QENS spectrometer used in the measurement. Therefore, only the diffusive components will contribute to the signal in QENS data - jump diffusion and localized caged diffusion. While they manifest as a sum of two exponentials in $I(Q,t)$, as shown in eq. \eqref{iisf acetamide final}, they will be observed as a sum of two Lorentzians in $S_{inc}(Q,E)$ which is measured in QENS experiments. This is clearly vindicated by the robustness of the two Lorentzian fit for the QENS data shown in Fig. \ref{qens des} (b). Clearly, jump diffusion component is a substantially narrow in comparison to the localized component, indicating that the jump-diffusion process occurs at a much slower timescale compared to localized caged diffusion.

\begin{figure}
    \centering
    \includegraphics[width=0.65\linewidth]{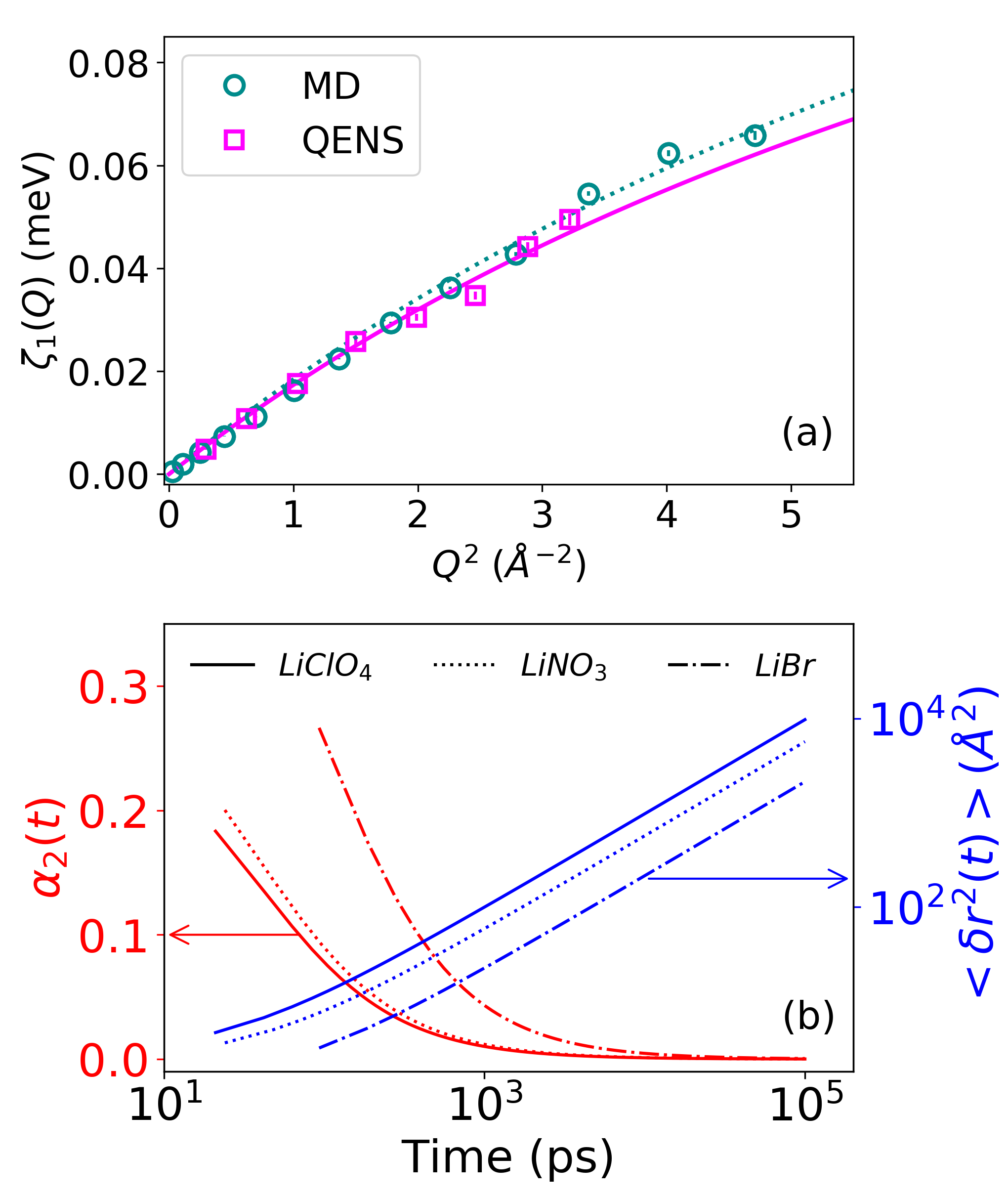}
    \caption{(a) The variation of inverse relaxation time associated to jump-diffusion, $\zeta_1$ with respect to $Q^2$ at a temperature of 365 K, as obtained from modelling MD simulation and QENS data. The corresponding lines indicate the model fitting based on NLD equation using an exponential kernel, as described in the text. (b) The time-dependent values of non-Gaussian parameters, $\alpha_2(t)$ (left-axis, red) and MSD, $\left< \delta r^2 (t)\right>$ (right axis, blue) for the jump diffusion process for three different DESs, as calculated from QENS experimental data at temperature of 330 K. Figure adapted from ref. \cite{Srinivasan_2023} }
    \label{hwhm msd}
\end{figure}

\subsubsection{Solutions from NLD for jump-diffusion}

To elucidate the jump-diffusion process in this system, we extend the non-local diffusion (NLD) equation to three dimensions (3D), considering the 3D version of equation \eqref{non-local diffusion} with a radially symmetric 3D kernel $\Lambda_h(\mathbf{r})$ due to the isotropy condition in the diffusion process. Under this condition, analogous to the rules provided in equation \eqref{limit_jump_rules}, we establish conditions for $\hat\Lambda_h(Q)$, where $Q = |\mathbf{Q}|$ represents the radial component of the momentum transfer. Utilizing a specific choice of $c_n = \left(-1\right)^{n-1}$ similar to equation \eqref{general jump-kernel}, we formulate the 3D kernel,
\begin{equation}
    \label{3D kernel}
    \Lambda_h(Q) = \frac{1}{\tau_j}\sum_{n=1}^\infty (-1)^n \left( Q l_0 \right)^{2(n-1)} = \frac{1}{\tau_j} \frac{l_0^2}{1 + \left( Q l_0 \right)^2 }
\end{equation}
where $l_0$ is the characteristic jump-length in the diffusion process and $\tau_j$ is the mean waiting time between jumps. If we define the jump-diffusivity, $D_j = l_0^2/\tau_j$, then the jump-kernel can be rewritten as $\Lambda_h(Q) = D_j\left(1 + D_j Q^2 \tau_j\right)^{-1}$. Inverting the Fourier-transform in 3D, we can find that the associated jump-kernel is,
\begin{equation}
    \label{3D exp kernel}
    \Lambda_h(r) = \frac{1}{4\pi\tau_j} \frac{\exp\left[-r/r_0\right]}{r}
\end{equation}
which has asymptotically exponential behaviour in $r$ and exhibits radial symmetry.

To independently probe jump-diffusion, we investigate the behavior of $I(Q,t)$ over substantially longer timescales, satisfying the condition $\zeta_2 t \ll 1$, which ensures probing the system beyond the caging timescale. At this timescale, contributions from localized diffusion and ballistic motion can be neglected, with only the jump-diffusion process being predominant. Consequently, equation \eqref{iisf acetamide final} simplifies to $I(Q,t) = C_0(Q) e^{-\zeta_1(Q) t}$. By comparing this with the solutions of the non-local diffusion (NLD) model in Fourier space (as given in equation \eqref{non-local diffusion FT}), we deduce that $\zeta_1(Q) = Q^2 \Lambda_h(Q)$, where $\Lambda_h(Q)$ represents the Fourier transform of the 3D jump kernel obtained in equation \eqref{3D kernel}. Therefore, we have,
\begin{equation}
    \label{hwhw jump}
    \zeta_1(Q) = \frac{D_j Q^2}{1 + D_j Q^2 \tau_j}
\end{equation}
where $D_j$ is the jump-diffusivity and $\tau_j$ is the mean residence time between jumps. Figure \ref{hwhm msd}(a) illustrates the inverse relaxation rate associated with the jump-diffusion process, $\zeta_1(Q)$, as a function of $Q^2$, obtained from both QENS and MD simulations \cite{Srinivasan_2020}. The fits based on equation \eqref{hwhw jump} for each dataset are also depicted in Figure \ref{hwhm msd}(a), affirming the effectiveness of the non-local diffusion (NLD) model in describing the jump-diffusion process in deep eutectic solvents (DESs).

Further, within the long-time approximation ($\zeta_2 t \ll 1$), it is also possible to extract the various moments using $I(Q,t)$ and the formula given in eq. \eqref{moments}. Using this, it can be shown that 
\begin{equation}
\begin{split}
    \label{msd ng jump}
    \left< \delta r^2(t) \right> = 6 \left( a + D_j t \right) \\
    \alpha_2(t) = \frac{1}{2} \frac{b + \tau_j D_j^2 t}{\left(a + D_j t \right)^2}
\end{split}
\end{equation}
 where $a$ and $b$ are related to statistical estimates of the cage sizes and can be estimated by modelling $C_0(Q)$ in the SISF. In Figure \ref{hwhm msd}(b), plots of equation \eqref{msd ng jump} are displayed for all three DESs, computed from fit parameters derived from QENS experiments. These plots clearly indicate a non-zero non-Gaussianity at long times, particularly characterized by the tail behavior $\alpha_2(t) \sim (\tau_j/t)$. Among the three systems, $\text{LiClO}_4$ exhibits the least heterogeneity and LiBr is the most heterogeneous. Notably, in all cases, the mean square displacement (MSD) exhibits linear time dependence, while the non-Gaussian parameter $\alpha_2(t)$ remains non-zero, suggesting the possibility of Fickian yet non-Gaussian diffusion. In the concluding section of this article, we will briefly remark the possibility of using NLD equation to precisely calculate solutions for Fickian yet non-Gaussian diffusion processes.

\section{Non-Gaussian fractional Brownian motion}
In the last two sections, we have discussed models that exhibit non-Markovian and non-Gaussian behaviour independently. While in the former, we chose the mathematical framework of SDEs to achieve non-Markovianity, the latter was accomplished by modifying the FPE for diffusion by incorporating non-locality. In this section, we aim to bring these two aspects together, by constructing a model which is simultaneously non-Markovian and non-Gaussian. In order to achieve this, we firstly look at the definition of fractional Brownian motion (fBm) process, $B_H(t)$, introduced by Mandelbrot and van Ness \cite{Mandelbrot_1968}
\begin{equation}
    \label{fBm definition}
    B_H(t) = \frac{1}{\Gamma(H+1/2)} \int\displaylimits_0^\infty (t-s)^{H-1/2} dW(s)
\end{equation}
where $H$ is called Hurst index and $W(t)$ is the Wiener's process defined according to the dimensionless version of the integral in eq. \eqref{Wiener_process}. This process reduces to the usual dimensionless Brownian motion (or Wiener's process) for the case $H = \frac{1}{2}$. Further, they represent a class of systems exhibiting subdiffusion for $H < 1/2$ and superdiffusion for $H>1/2$. In studying non-Markovian processes in molecular fluids, we are particularly interested in systems exhibiting subdiffusion. The general statistical properties of the fBm follow,
\begin{equation}
\begin{split}
    \label{fBm properties}
    \left< B_H(t) \right> = 0  \\
    \left< B_H(t_1) B_H(t_2) \right> = \left( t_1^{2H} + t_2^{2H} - |t_1 - t_2|^{2H} \right)
\end{split}
\end{equation}
The dispalcement of the particle undergoing subdiffusion can be described on the basis of increments of the fBm process, such that $dx(t) = \sqrt{2D_H} dB_H(t)$, where $D_H$ is dimensional factor which holds a physical resemblance to diffusivity. In such a scenario, it is possible to obtain a FPE for the displacement of the particle using the recent developments of Stochastic calculus of fBm processes \cite{Duncan_2000}. To this end, let us consider an associated stochastic process, $h = h(x(t))$, then according to Theorem 4.2 in \cite{Duncan_2000}, we have the relationship,
\begin{equation}
    \label{fbm step1}
    \frac{dh(x)}{dt} = \sqrt{2D_H} \left[ \frac{d^2h}{dx^2} \int\displaylimits_0^\infty \sqrt{2D_H}\phi(t,s)ds + \frac{dh}{dx} \frac{dB_H(t)}{dt}\right]
\end{equation}
where $\phi(s,t)=H(2H-1)|s-t|^{2H-2}$. To derive the FPE for the displacement, we compute the expectation value of the above equation,
\begin{equation}
    \label{fbm step2}
    \int\displaylimits_{-\infty}^{\infty}dx \frac{\partial P(x,t)}{\partial t}h(x) = 2D_H Ht^{2H-1}\int\displaylimits_{-\infty}^{\infty}dx \frac{\partial^2 P(x,t)}{\partial x^2}h(x)
\end{equation}
where we have used $\left< dB_H(t)\right>=0$ to evaluate the expectation value of the last term in eq. \eqref{fbm step1}. Since the eq. \eqref{fbm step2} is valid for any arbitrary function, $h(x)$, we have,
\begin{equation}
    \label{fBm FPE}
    \frac{\partial P(x,t)}{\partial t} = \alpha t^{\alpha-1}D_{\alpha}\frac{\partial^2 P(x,t)}{\partial x^2}
\end{equation}
where we have considered $\alpha = 2H$ and  $D_\alpha = D_H$ for convenience. The eq. \eqref{fBm FPE} is the FPE for the displacement of the particle following a fBm process subject to open boundary conditions, i.e. $x\in(-\infty,\infty)$. A straightforward calculation yields the behaviour of MSD, $\left< x^2(t) \right> = 2D_\alpha t^\alpha$. Further, subject to the initial condition $P(x,0) = \delta(x)$, we also note that the solution of eq. \eqref{fBm FPE} is Gaussian in $x$,
\begin{equation}
    \label{fBm FPE soln}
    P(x,t) = \frac{1}{\sqrt{4\pi D_\alpha t^\alpha}}\exp\left[ \frac{x^2}{4D_\alpha t^\alpha}\right]
\end{equation}
The SISF of the fBm exhibits a stretched exponential, $I(k,t) = \exp\left[-D_\alpha k^2 t^\alpha\right]$. The nonexponentiality of the SISF is a key signature of the non-Markovian nature of the fBm process. However, it's important to note that fBm process remains inherently Gaussian. In fact, it is connected to the long-time behavior of the diffusion process described using a GLE with a power-law memory kernel. In order to understand this connection, we rewrite the definition of fBm in terms of fractional Gaussian noise (fGn) \cite{Kou_2004}
\begin{equation}
    \label{fBm fGn}
    B_H(t) =\frac{1}{\Gamma(H+1/2)}\int\displaylimits_0^t (t-s)^{H-1/2}\zeta_H (s) ds
\end{equation}
where $\zeta_H (t) = \frac{dB_H(t)}{dt}$ is the fractional Gaussian noise (fGn) process. Although fGn bears resemblance to Gaussian white noise, it has a power-law decaying autocorrelation of the form $\left< \zeta_H(t)\zeta_H(t') \right> = 2H(H-1) |t-t'|^{2H-2} + 4H|t-t'|^{2H-1}\delta(t-t')$ instead of a delta-correlation found in white noise. Therefore, considering fGn, $\zeta_H(t)$, in the GLE (eq. \eqref{GLE}), fixes the memory kernel to follow $M(t) \sim t^{2H-2}$ which is akin to the power-law memory kernel which was discussed in section 2.2.

While fBm serves as a robust model to describe the Gaussian subdiffusion, we now turn to construct a model that can explain non-Gaussian subdiffusion. This involves introducing a jump-kernel and redefing the  FPE in eq. \eqref{fBm FPE} as a non-local equation for non-Gaussian fractional Brownian motion (nGfBm) \cite{Srinivasan_2024},

\begin{equation}
    \label{nGfBm FPE}
    \frac{\partial P(x,t)}{\partial t} = \alpha t^{\alpha-1}D_{\alpha}\int\displaylimits_{-\infty}^\infty \Lambda_h(x-x') \frac{\partial^2 P(x',t)}{\partial x'^2} dx'
\end{equation}
where $\Lambda_h(x)$ is the jump-kernel as defined in the previous section. It is easy to verify that choosing $\Lambda_h(x) = D_\alpha \delta(x)$ leads to fBm which is essentially a Gaussian process. Yet again, the solutions to eq. \eqref{nGfBm FPE} is easily obtained in the Fourier domain with SISF being given as a stretched exponential,
\begin{equation}
    \label{nGfBm iqt}
    I(k,t) = I_0(k) \exp\left[-k^2 \hat\Lambda_h(k) t^\alpha\right]
\end{equation}
The various possible jump-kernels that have been described in the previous section can be employed for the solutions in nGfBm too, wherein the system will exhibit a strongly subdiffusive dynamics but transition from non-Gaussian at small length scales to Gaussian behaviour at large distances. However, it should be noted that the limiting behaviour of $\hat\Lambda_h(k)$ will be rescaled over $\tau_j^\alpha$ according to,
\begin{equation}
    \label{limit rules fBm}
    \hat\Lambda_h(k) \xrightarrow{k\rightarrow 0} \frac{x_0^2}{\tau_j^\alpha} \propto D_\alpha \qquad \qquad
        \hat\Lambda_h(k) \xrightarrow{k\rightarrow\infty} \frac{k^{-2}}{\tau_j^\alpha}
\end{equation}
to obtain the appropriate limiting conditions in the Gaussian ($k\rightarrow 0$) and non-Gaussian ($k\rightarrow \infty$) limits. Respecting these conditions, the jump-kernel $\hat\Lambda_h(k)$ listed in Table 1.1 will also be defined with a scaling time of $\tau_j^{-\alpha}$ instead of $\tau_j^{-1}$. 

The most remarkable feature of the nGfBm model lies in its ability to capture both the non-Gaussian and non-Markovian characteristics of the diffusion process. It's essential to recognize that the Fokker-Planck equation (FPE) for nGfBm, as defined in Equation \eqref{fBm FPE}, cannot be directly derived from a continuous time random walk approach due to its inherently non-Markovian nature. As a result, eq. \eqref{fBm FPE} along with the limiting conditions in eq. \eqref{limit rules fBm} represents a unique approach for understanding diffusion mechanisms that involve a transition from non-Gaussian to Gaussian subdiffusion, a transition fundamentally driven by jump diffusion with long-term temporal correlations. On the other hand, relaxing the conditions given in eq. \eqref{limit rules fBm} on the jump-kernel, we can establish methods to treat completely non-Gaussian diffusion processes with inherent long-term temporal correlations. A typical example of that kind would be the treatment of Levy flights driven by fBm, which shall be explored in future. 

\subsection{Subdiffusion crossover in glass-formers}
Molecular and polymeric glass formers commonly exhibit a crossover from non-Gaussian to Gaussian diffusion while still being non-Fickian (subdiffusive) \cite{Srinivasan_2024, Busselez_2011,Kofu_2018,Arbe_2002,Arbe_2003,Capponi_2009,Arbe_1998}. Although extensively studied via simulations \cite{Arbe_2002, Busselez_2011,Colmenero_2002} and experiments \cite{Arbe_2002,Busselez_2011,Kofu_2018,Arbe_2003,Capponi_2009,Colmenero_2002}, a comprehensive understanding of the underlying basis for non-Gaussianity across these crossovers remains elusive due to the absence of a first-principles model. Glass formers typically exhibit an exponential decay in their displacement distribution \cite{Wang_2009,Chubynsky_2014,Chaudhuri_2007}, a characteristic observed in various complex fluids such as colloidal suspensions \cite{Rusciano_2022, Pastore_2021,Wang_2009}, Si atoms in a silica melt \cite{Chaudhuri_2007,Berthier_2007}, and Lennard-Jones particles \cite{Chaudhuri_2007, Berthier_2007a}. This behavior is attributed to large deviations and randomization of the number of jumps in particle displacement \cite{Barkai_2020}. However, the precise nature of the displacement distribution in glass formers undergoing subdiffusion crossover remains unexplored. 

In glass-formers the SISF is typically found to follow a stretched exponential function, $I_s (Q,t)=\exp \left[-{(t/\tau_s (Q))}^\beta \right]$. Here, $\beta$ represents the stretching parameter, reflecting the departure from an exponential relaxation profile, while $\tau_s$ denotes the characteristic relaxation time. Numerous experimental \cite{Arbe_2002, Kofu_2018, Colmenero_2002, Capponi_2009, Arbe_1998} and computational\cite{Busselez_2011, Colmenero_2002, Capponi_2009} investigations reveal a crossover in $\tau_s$ vs $Q$ relationship near the first maximum, $Q_0$ of the structure factor. Essentially, for $Q < Q_0$, $\tau(Q) \sim Q^{-2/\beta}$, meanwhile for $Q > Q_0$, $\tau(Q) \sim Q^{-2}$. At low-Q values ($< Q_0$), juxtaposing the relationship $\tau_s  \sim Q^{-2/\beta}$ with the stretched exponential decay, inevitably leads to a Gaussian sub-diffusion with a MSD behaviour, $\left<\delta r^2 (t)\right> \sim t^\beta$. But, in the high $Q$ regime ($> Q_0$), where $\tau_s  \sim Q^{-2}$, the diffusion mechanism cannot be described within the Gaussian approximation \cite{Arbe_2002, Busselez_2011, Arbe_1998}. Therefore, this crossover in the behaviour of relaxation time, has been attributed to a transition from Gaussian (for $Q < Q_0$) to a non-Gaussian (for $Q > Q_0$) sub-diffusion in these media\cite{Arbe_2002, Busselez_2011, Kofu_2018}. In the subsequent discussion, we delve into characterizing this sub-diffusion crossover by formulating a Fokker-Planck equation for the nGfBm model. Our findings underscore that the emergence of this crossover is exclusively a result of the non-locality induced by the jump-kernel in the model.

\subsubsection{Modelling QENS data in glass-formers}

\begin{figure}
    \centering
    \includegraphics[width=0.6\linewidth]{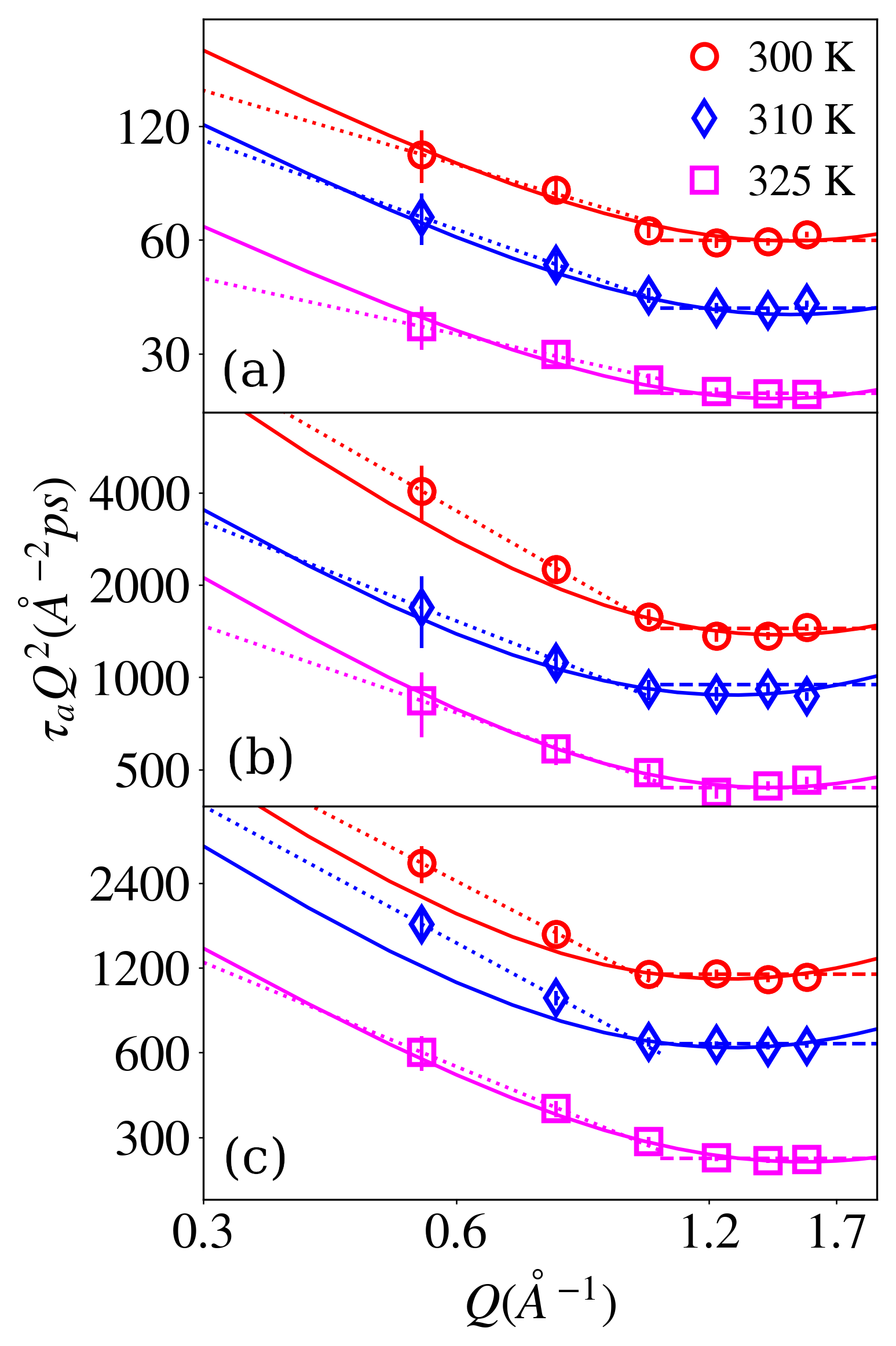}
    \caption{Plot of $\tau_a Q^2$ vs $Q$ for (a) ethylene glycol (EG), (b) EG + LiCl and (c) EG + $\text{ZnCl}_2$ obtained from the fits of IQENS experimental data. Also shown are the fits based on nGfBm model employing exponential kernel (Model A). }
    \label{tau 3 plots}
\end{figure}

The QENS data for ethylene glycol (EG) and its associated DESs like EG + $\text{ZnCl}_2$ (1:4 molar ratio) and EG + LiCl (1:3 molar ratio) exhibit a stretched exponential relaxation profile, similar to what has been observed in various other glass-forming systems \cite{Arbe_1998, Arbe_2002, Capponi_2009, Kofu_2018, Busselez_2011}. This profile is characterized by a characteristic timescale $\tau_s$ and a stretching exponent $\beta$. Liquid EG is known to exhibit stretched exponential relaxation \cite{Crupi_2003, Sobolev_2007}, and it is anticipated that DESs based on EG will also display this characteristic, given their resemblance to supercooled liquids\cite{Banerjee_2020}. The average relaxation time $\tau_a(Q) = \tau_s(Q)\beta^{-1}\Gamma(\beta^{-1})$ is calculated from QENS data fitting. Fig. \ref{tau 3 plots} shows the variation of $\tau_a$ vs $Q$ for all three systems (EG, EG+$\text{ZnCl}_2$, EG+LiCl). As illustrated in the plots, these systems show a crossover from $Q^{-2/\beta}$ (for $Q < 1 \text{\AA}^{-1}$, dotted lines) to $Q^{-2}$ (for $Q > 1 \text{\AA}^{-1}$, dashed lines). This clearly indicates the crossover from Gaussian dynamics at low $Q$ to non-Gaussian behaviour at higher $Q$-values for EG and the DESs. 

\subsubsection{Employing nGfBm for QENS data}
In order to apply the framework of nGfBm, we extend the eq. \eqref{nGfBm FPE} to 3D systems with radial symmetry, with jump-kernel defined through the function $\Lambda_h(r)$. Therefore, SISF for the 3D nGfBm process can be given as,
\begin{equation}
    \label{3d nGfBm}
    I_s(Q,t) = I_0(Q) \exp \left[- Q^2 \hat\Lambda_h(Q) t^\beta \right]
\end{equation}
where $\hat\Lambda_h(Q) $ is the 3D Fourier transform of $\Lambda_h(r)$. Upon comparison, an expression for the variation of average relaxation time with respect to $Q$ can be given, $\tau_a(Q) = (Q^2 \hat\Lambda_h(Q))^{-1/\beta}$. With conditions similar to that expressed in eq. \eqref{limit rules fBm}, we can clearly show the limiting behaviour of $\tau_a(Q)$
\begin{equation}
    \label{limit fBm tau}
    \tau_a(Q) \xrightarrow{Q\rightarrow 0} D_{\beta}^{-1/\beta} Q^{-2/\beta} ;\qquad
        \tau_a(Q) \xrightarrow{Q\rightarrow\infty} \tau_j
\end{equation}
where $D_\beta = l_0^2/\tau_j^\beta$, wherein $l_0$ is characteristic jump-length. Hence, the nGfBm model inherently encompasses the transition from non-Gaussian to Gaussian subdiffusion behavior. Crucially, the $Q$-value marking this crossover is determined by the characteristic jump-length, $l_0$. While various options are available for the jump-kernel to reproduce the limiting behavior described in equation \eqref{limit fBm tau}, we opt for the exponential kernel, where $c_n = (-1)^{n-1}$. As shown in eq. \eqref{3D exp kernel}, this amounts to choosing an radially symmetric exponential kernel. In this case, the behaviour of the average relaxation time is $\tau_a(Q) = \tau_j \left[1 + (Ql_0)^{-2} \right]^{1/\beta}$. The model fits based on this expression is shown in Fig. \ref{tau 3 plots} as solid lines. It is evident that the model describes the data really well and smoothly exhibits the crossover from Gaussian to non-Gaussian behaviour. To be more precise, the point of crossover can be exactly calculated to be $Q^* = \sqrt{(1/\beta) -1}/l_0$ ($\beta\ne 1$). It is observed that the values of $Q^*$ is between $ 1 - 2 \;\text{\AA}^{-1}$ for all the three systems, as the value of $l_0$ falls around $\sim 0.6 \; \text{\AA}$. Therefore, it is evident that the crossover point in $Q$-space is controlled by $l_0$ suggesting an inverse correlation between them.   

\begin{figure}
    \centering
    \includegraphics[width=0.6\linewidth]{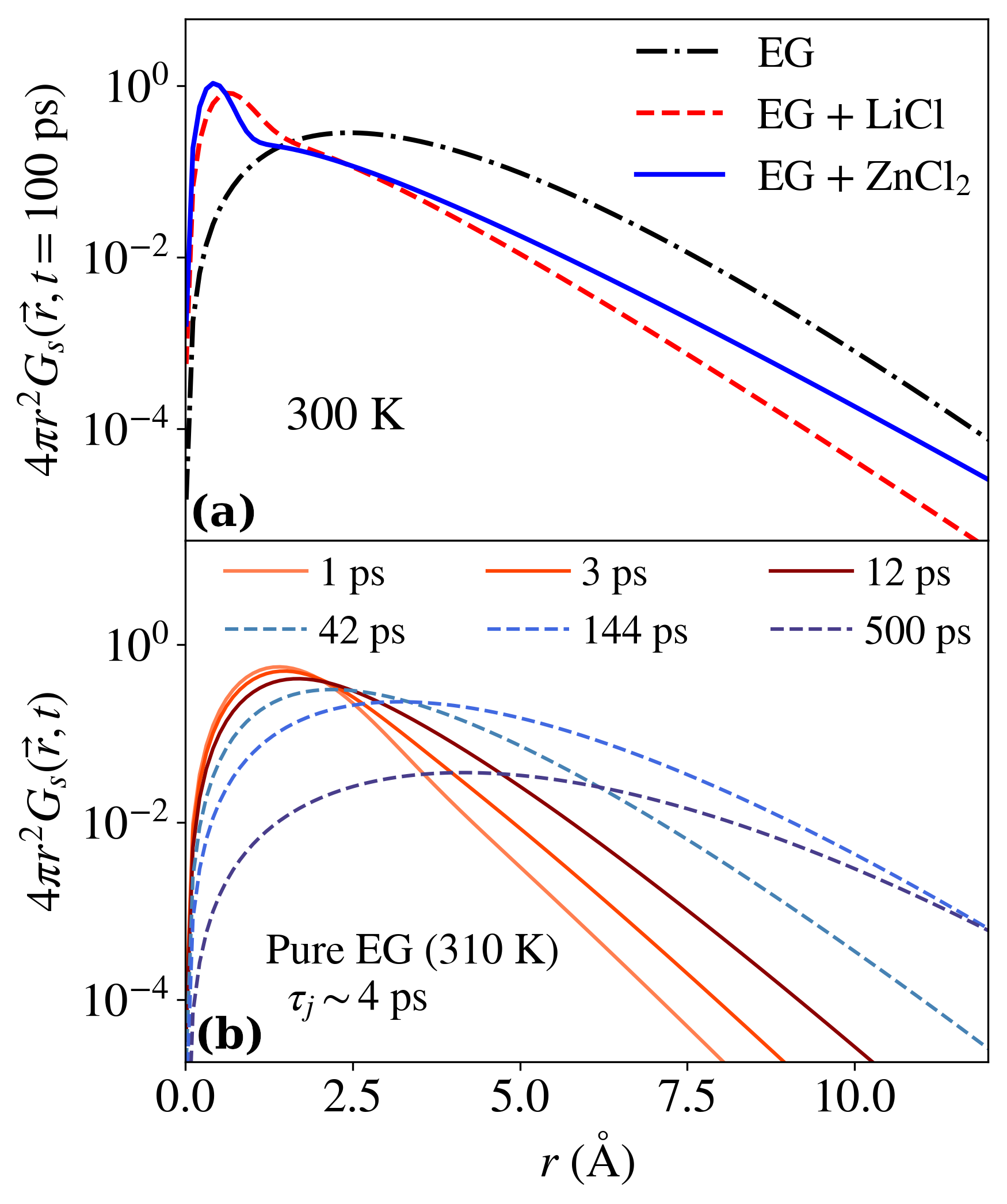}
    \caption{(a) The radial van-Hove self-correlation function calculated based on the parameters extracted from experimental QENS fits for pure EG and DESs (EG + LiCl, EG + $\text{ZnCl}_2$) at T = 300 K and $t$ = 100 ps. (b) Radial van-Hove self-correlation function for pure EG (310 K) calculated at different times $t$; $t > 10\tau_j$ are shown by broken lines and $t < 10\tau_j$ are shown by solid lines.  Figure adapted from ref. \cite{Srinivasan_2024}}
    \label{gsrt full}
\end{figure}

\subsubsection{Emergence of exponential tails in nGfBm}
In order to show that exponential tails emerge naturally in the nGfBm model, in Fig. \ref{gsrt full} we plot the radial van-Hove self-correlation function, $G_s(r,t)$ for all the systems pure EG and DESs (EG + LiCl, EG + $\text{ZnCl}_2$), as calculated from the model fits of QENS data. Figure \ref{gsrt full}(a) illustrates the plots of $4\pi r^2 G_s (\mathbf{r} ,t=100 ps)$ for both the DESs and pure EG at 300 K. The notable difference in the values of $\tau_j$ for EG and DESs, approximately 7 ps and 200 ps, respectively, explains the distinctiveness observed in their curves. In DESs, the longer $\tau_j$ leads to more pronounced initial peaks ($r < 2 \text{\AA}$) in the van-Hove self-correlation function. Conversely, these peaks vanish in pure EG, where the fast local dynamics have fully relaxed within 100 ps, rendering the dynamics purely diffusive. Additionally, while the tails prominently display an exponential decay in both DESs, pure EG exhibits a nearly Gaussian behavior. A clearer insight into these changes is provided by examining the plots of the radial van-Hove self-correlation for pure EG (310 K) at various times (ranging from 1 to 500 ps) in Figure \ref{gsrt full}(b). Notably, the presence of exponential tails is discernible for $t < 10 \tau_j$. To emphasize this characteristic and illustrate the transition from the non-Gaussian to Gaussian regime, solid lines represent curves for $t < 10 \tau_j$, while dashed lines denote those for $t > 10 \tau_j$.

\section{Concluding Remarks}
The Brownian motion model has played a crucial role in advancing our understanding across diverse disciplines such as biology, materials science, finance, and environmental science. By providing insights into stochastic processes, it has facilitated numerous discoveries and applications. Yet, as experimental and simulation techniques evolve, we increasingly uncover instances where the assumptions of Gaussianity and Markovianity inherent in the Brownian motion model are inadequate. This realization underscores the necessity for alternative models that can capture the complexities of real-world systems without relying on these assumptions. This perspective article highlights recent advancements in addressing challenges encountered in modeling the diffusion mechanism of molecules in complex fluids. Initially, the approach focuses on studying systems that manifest either non-Gaussian or non-Markovian behavior exclusively. We consider mathematical models that extend beyond the traditional Brownian regime, offering a robust framework for characterizing diffusion in such systems. These models are strongly supported by experimental studies utilizing neutron scattering and computational studies employing MD simulations.

The Generalized Langevin equation (GLE) provides a framework for describing systems with non-Markovian dynamics driven by colored noise, such as power-law noise. This extension allows for modeling fractional Brownian motion (fBm), which deviates notably from traditional Brownian motion. In studies of lipid lateral motion, fractional Gaussian noise (fGn) has been identified as a driving force, characterized by a power-law memory kernel within the GLE framework. Molecular dynamics (MD) simulations further confirm the lateral diffusion as a Gaussian process, supporting the use of fGn. Experimental findings from Quasi-Elastic Neutron Scattering (QENS) experiments also exhibit stretched exponential relaxation behavior, aligning with this description.

On the other hand, the non-local diffusion (NLD) model, with its non-local jump kernels, inherently exhibits non-Gaussianity, resulting in spatially heterogeneous diffusion processes. This trait is validated through the calculation of the non-Gaussian parameter in MD simulations. It's thereby established that the cage-jump diffusion mechanism inherently introduces dynamical heterogeneity in the system. This article provides evidence of applicability of NLD in explaining the diffusion mechanism of molecules in deep eutectic solvents (DESs), paving way to explain the origin of dynamical heterogeneity in these complex media.

The previous section delves into systems exhibiting simultaneous non-Markovianity and non-Gaussianity. In such cases, traditional Markovian continuous time random walk (CTRW) models fall short due to jumps with inherent long time history dependence. Introducing the non-Gaussian fractional Brownian motion (nGfBm) model addresses this limitation, offering a modified version of NLD for systems driven by colored noise. The nGfBm model provides a valuable framework for naturally invoking the emergence of sub-diffusion crossover in glass-formers. This model also provides a method to show the origin of exponential tails in such dynamically heterogeneous systems.

This perspective article provides a comprehensive overview of molecular diffusion which exhibit violation of Brownian motion. These violations have been described using frameworks which essentially incorporate non-Markovianity and non-Gaussianity into the diffusion phenomena. Additionally, these models have also been applied and validated on data obtained on various systems using QENS experiments and MD simulations.

\bibliography{rsc} 

\providecommand*{\mcitethebibliography}{\thebibliography}
\csname @ifundefined\endcsname{endmcitethebibliography}
{\let\endmcitethebibliography\endthebibliography}{}
\begin{mcitethebibliography}{70}
\providecommand*{\natexlab}[1]{#1}
\providecommand*{\mciteSetBstSublistMode}[1]{}
\providecommand*{\mciteSetBstMaxWidthForm}[2]{}
\providecommand*{\mciteBstWouldAddEndPuncttrue}
  {\def\EndOfBibitem{\unskip.}}
\providecommand*{\mciteBstWouldAddEndPunctfalse}
  {\let\EndOfBibitem\relax}
\providecommand*{\mciteSetBstMidEndSepPunct}[3]{}
\providecommand*{\mciteSetBstSublistLabelBeginEnd}[3]{}
\providecommand*{\EndOfBibitem}{}
\mciteSetBstSublistMode{f}
\mciteSetBstMaxWidthForm{subitem}
{(\emph{\alph{mcitesubitemcount}})}
\mciteSetBstSublistLabelBeginEnd{\mcitemaxwidthsubitemform\space}
{\relax}{\relax}

\bibitem[Einstein(1905)]{Einstein_1905}
A.~Einstein, \emph{Annalen der Physik}, 1905, \textbf{322}, 549--560\relax
\mciteBstWouldAddEndPuncttrue
\mciteSetBstMidEndSepPunct{\mcitedefaultmidpunct}
{\mcitedefaultendpunct}{\mcitedefaultseppunct}\relax
\EndOfBibitem
\bibitem[Zwanzig(2001)]{zwanzig2001nonequilibrium}
R.~Zwanzig, \emph{Nonequilibrium Statistical Mechanics}, Oxford University
  Press, 2001\relax
\mciteBstWouldAddEndPuncttrue
\mciteSetBstMidEndSepPunct{\mcitedefaultmidpunct}
{\mcitedefaultendpunct}{\mcitedefaultseppunct}\relax
\EndOfBibitem
\bibitem[Langevin(1908)]{Langevin_1908}
P.~Langevin, \emph{CR Acad. Sci. Paris}, 1908, \textbf{146}, 530\relax
\mciteBstWouldAddEndPuncttrue
\mciteSetBstMidEndSepPunct{\mcitedefaultmidpunct}
{\mcitedefaultendpunct}{\mcitedefaultseppunct}\relax
\EndOfBibitem
\bibitem[Gardiner(2009)]{gardiner2009}
C.~Gardiner, \emph{Stochastic methods}, "Springer", 2009, vol.~4\relax
\mciteBstWouldAddEndPuncttrue
\mciteSetBstMidEndSepPunct{\mcitedefaultmidpunct}
{\mcitedefaultendpunct}{\mcitedefaultseppunct}\relax
\EndOfBibitem
\bibitem[B{\'e}e(1988)]{Bee_1988}
M.~B{\'e}e, \emph{Quasielastic Neutron Scattering : Principles and Applications
  in Solid State Chemistry, Biology and Materials Science}, {A. Hilger},
  {Bristol; Philadelphia}, 1988\relax
\mciteBstWouldAddEndPuncttrue
\mciteSetBstMidEndSepPunct{\mcitedefaultmidpunct}
{\mcitedefaultendpunct}{\mcitedefaultseppunct}\relax
\EndOfBibitem
\bibitem[Arbe \emph{et~al.}(1998)Arbe, Colmenero, Monkenbusch, and
  Richter]{Arbe_1998}
A.~Arbe, J.~Colmenero, M.~Monkenbusch and D.~Richter, \emph{Physical Review
  Letters}, 1998, \textbf{81}, 590--593\relax
\mciteBstWouldAddEndPuncttrue
\mciteSetBstMidEndSepPunct{\mcitedefaultmidpunct}
{\mcitedefaultendpunct}{\mcitedefaultseppunct}\relax
\EndOfBibitem
\bibitem[Arbe \emph{et~al.}(2002)Arbe, Colmenero, Alvarez, Monkenbusch,
  Richter, Farago, and Frick]{Arbe_2002}
A.~Arbe, J.~Colmenero, F.~Alvarez, M.~Monkenbusch, D.~Richter, B.~Farago and
  B.~Frick, \emph{Physical Review Letters}, 2002, \textbf{89}, 245701\relax
\mciteBstWouldAddEndPuncttrue
\mciteSetBstMidEndSepPunct{\mcitedefaultmidpunct}
{\mcitedefaultendpunct}{\mcitedefaultseppunct}\relax
\EndOfBibitem
\bibitem[Burankova \emph{et~al.}(2014)Burankova, Hempelmann, Wildes, and
  Embs]{Burankova_2014}
T.~Burankova, R.~Hempelmann, A.~Wildes and J.~P. Embs, \emph{The Journal of
  Physical Chemistry B}, 2014, \textbf{118}, 14452--14460\relax
\mciteBstWouldAddEndPuncttrue
\mciteSetBstMidEndSepPunct{\mcitedefaultmidpunct}
{\mcitedefaultendpunct}{\mcitedefaultseppunct}\relax
\EndOfBibitem
\bibitem[Busch \emph{et~al.}(2010)Busch, Smuda, Pardo, and Unruh]{busch_2010}
S.~Busch, C.~Smuda, L.~C. Pardo and T.~Unruh, \emph{Journal of the American
  Chemical Society}, 2010, \textbf{132}, 3232--3233\relax
\mciteBstWouldAddEndPuncttrue
\mciteSetBstMidEndSepPunct{\mcitedefaultmidpunct}
{\mcitedefaultendpunct}{\mcitedefaultseppunct}\relax
\EndOfBibitem
\bibitem[Sharma \emph{et~al.}(2010)Sharma, Mitra, Verma, Hassan, Garcia~Sakai,
  and Mukhopadhyay]{Sharma_2010}
V.~K. Sharma, S.~Mitra, G.~Verma, P.~A. Hassan, V.~Garcia~Sakai and
  R.~Mukhopadhyay, \emph{The Journal of Physical Chemistry B}, 2010,
  \textbf{114}, 17049--17056\relax
\mciteBstWouldAddEndPuncttrue
\mciteSetBstMidEndSepPunct{\mcitedefaultmidpunct}
{\mcitedefaultendpunct}{\mcitedefaultseppunct}\relax
\EndOfBibitem
\bibitem[Sharma \emph{et~al.}(2012)Sharma, Mitra, Sakai, Hassan, Embs, and
  Mukhopadhyay]{Sharma_2012}
V.~K. Sharma, S.~Mitra, V.~G. Sakai, P.~A. Hassan, J.~P. Embs and
  R.~Mukhopadhyay, \emph{Soft Matter}, 2012, \textbf{8}, 7151--7160\relax
\mciteBstWouldAddEndPuncttrue
\mciteSetBstMidEndSepPunct{\mcitedefaultmidpunct}
{\mcitedefaultendpunct}{\mcitedefaultseppunct}\relax
\EndOfBibitem
\bibitem[Srinivasan \emph{et~al.}(2018)Srinivasan, Sharma, Mitra, and
  Mukhopadhyay]{Srinivasan_2018}
H.~Srinivasan, V.~K. Sharma, S.~Mitra and R.~Mukhopadhyay, \emph{Journal of
  Physical Chemistry C}, 2018,  20419–20430\relax
\mciteBstWouldAddEndPuncttrue
\mciteSetBstMidEndSepPunct{\mcitedefaultmidpunct}
{\mcitedefaultendpunct}{\mcitedefaultseppunct}\relax
\EndOfBibitem
\bibitem[Srinivasan \emph{et~al.}(2020)Srinivasan, Sharma, Sakai, Embs,
  Mukhopadhyay, and Mitra]{Srinivasan_2020}
H.~Srinivasan, V.~K. Sharma, V.~G. Sakai, J.~P. Embs, R.~Mukhopadhyay and
  S.~Mitra, \emph{The Journal of Physical Chemistry B}, 2020, \textbf{124},
  1509--1520\relax
\mciteBstWouldAddEndPuncttrue
\mciteSetBstMidEndSepPunct{\mcitedefaultmidpunct}
{\mcitedefaultendpunct}{\mcitedefaultseppunct}\relax
\EndOfBibitem
\bibitem[Aoun \emph{et~al.}(2015)Aoun, Sharma, Pellegrini, Mitra, Johnson, and
  Mukhopadhyay]{Aoun_2015}
B.~Aoun, V.~K. Sharma, E.~Pellegrini, S.~Mitra, M.~Johnson and R.~Mukhopadhyay,
  \emph{The Journal of Physical Chemistry B}, 2015, \textbf{119},
  5079--5086\relax
\mciteBstWouldAddEndPuncttrue
\mciteSetBstMidEndSepPunct{\mcitedefaultmidpunct}
{\mcitedefaultendpunct}{\mcitedefaultseppunct}\relax
\EndOfBibitem
\bibitem[Armstrong \emph{et~al.}(2011)Armstrong, Trapp, Peters, Seydel, and
  Rheinst{\"a}dter]{Armstrong_2011}
C.~L. Armstrong, M.~Trapp, J.~Peters, T.~Seydel and M.~C. Rheinst{\"a}dter,
  \emph{Soft Matter}, 2011, \textbf{7}, 8358--8362\relax
\mciteBstWouldAddEndPuncttrue
\mciteSetBstMidEndSepPunct{\mcitedefaultmidpunct}
{\mcitedefaultendpunct}{\mcitedefaultseppunct}\relax
\EndOfBibitem
\bibitem[Sharma \emph{et~al.}(2015)Sharma, Mamontov, Anunciado, O'Neill, and
  Urban]{Sharma_2015}
V.~K. Sharma, E.~Mamontov, D.~B. Anunciado, H.~O'Neill and V.~Urban, \emph{The
  Journal of Physical Chemistry B}, 2015, \textbf{119}, 4460--4470\relax
\mciteBstWouldAddEndPuncttrue
\mciteSetBstMidEndSepPunct{\mcitedefaultmidpunct}
{\mcitedefaultendpunct}{\mcitedefaultseppunct}\relax
\EndOfBibitem
\bibitem[Sharma \emph{et~al.}(2016)Sharma, Mamontov, Tyagi, Qian, Rai, and
  Urban]{Sharma_2016}
V.~K. Sharma, E.~Mamontov, M.~Tyagi, S.~Qian, D.~K. Rai and V.~S. Urban,
  \emph{The Journal of Physical Chemistry Letters}, 2016, \textbf{7},
  2394--2401\relax
\mciteBstWouldAddEndPuncttrue
\mciteSetBstMidEndSepPunct{\mcitedefaultmidpunct}
{\mcitedefaultendpunct}{\mcitedefaultseppunct}\relax
\EndOfBibitem
\bibitem[Dubey \emph{et~al.}(2018)Dubey, Srinivasan, Sharma, Mitra, Sakai, and
  Mukhopadhyay]{Dubey_2018}
P.~S. Dubey, H.~Srinivasan, V.~K. Sharma, S.~Mitra, V.~G. Sakai and
  R.~Mukhopadhyay, \emph{Scientific Reports}, 2018, \textbf{8}, 1862\relax
\mciteBstWouldAddEndPuncttrue
\mciteSetBstMidEndSepPunct{\mcitedefaultmidpunct}
{\mcitedefaultendpunct}{\mcitedefaultseppunct}\relax
\EndOfBibitem
\bibitem[Dueby \emph{et~al.}(2019)Dueby, Dubey, and Daschakraborty]{Dueby_2019}
S.~Dueby, V.~Dubey and S.~Daschakraborty, \emph{The Journal of Physical
  Chemistry B}, 2019, \textbf{123}, 7178--7189\relax
\mciteBstWouldAddEndPuncttrue
\mciteSetBstMidEndSepPunct{\mcitedefaultmidpunct}
{\mcitedefaultendpunct}{\mcitedefaultseppunct}\relax
\EndOfBibitem
\bibitem[Falck \emph{et~al.}(2008)Falck, R{\'o}g, Karttunen, and
  Vattulainen]{falck_2008}
E.~Falck, T.~R{\'o}g, M.~Karttunen and I.~Vattulainen, \emph{Journal of the
  American Chemical Society}, 2008, \textbf{130}, 44--45\relax
\mciteBstWouldAddEndPuncttrue
\mciteSetBstMidEndSepPunct{\mcitedefaultmidpunct}
{\mcitedefaultendpunct}{\mcitedefaultseppunct}\relax
\EndOfBibitem
\bibitem[Wanderlingh \emph{et~al.}(2014)Wanderlingh, D'Angelo, Branca,
  Conti~Nibali, Trimarchi, Rifici, Finocchiaro, Crupi, Ollivier, and
  Middendorf]{wanderlingh_2014}
U.~Wanderlingh, G.~D'Angelo, C.~Branca, V.~Conti~Nibali, A.~Trimarchi,
  S.~Rifici, D.~Finocchiaro, C.~Crupi, J.~Ollivier and H.~D. Middendorf,
  \emph{The Journal of Chemical Physics}, 2014, \textbf{140}, 174901\relax
\mciteBstWouldAddEndPuncttrue
\mciteSetBstMidEndSepPunct{\mcitedefaultmidpunct}
{\mcitedefaultendpunct}{\mcitedefaultseppunct}\relax
\EndOfBibitem
\bibitem[Flenner \emph{et~al.}(2009)Flenner, Das, Rheinst{\"a}dter, and
  Kosztin]{Flenner_2009}
E.~Flenner, J.~Das, M.~C. Rheinst{\"a}dter and I.~Kosztin, \emph{Physical
  Review E}, 2009, \textbf{79}, 011907\relax
\mciteBstWouldAddEndPuncttrue
\mciteSetBstMidEndSepPunct{\mcitedefaultmidpunct}
{\mcitedefaultendpunct}{\mcitedefaultseppunct}\relax
\EndOfBibitem
\bibitem[Jeon \emph{et~al.}(2012)Jeon, Monne, Javanainen, and
  Metzler]{Jeon_2012}
J.-H. Jeon, H.~M.-S. Monne, M.~Javanainen and R.~Metzler, \emph{Physical Review
  Letters}, 2012, \textbf{109}, 188103\relax
\mciteBstWouldAddEndPuncttrue
\mciteSetBstMidEndSepPunct{\mcitedefaultmidpunct}
{\mcitedefaultendpunct}{\mcitedefaultseppunct}\relax
\EndOfBibitem
\bibitem[Sahu \emph{et~al.}(2023)Sahu, Srinivasan, Jadhav, Sharma, and
  Debnath]{Samapika_2023}
S.~Sahu, H.~Srinivasan, S.~E. Jadhav, V.~K. Sharma and A.~Debnath,
  \emph{Langmuir}, 2023, \textbf{39}, 16432--16443\relax
\mciteBstWouldAddEndPuncttrue
\mciteSetBstMidEndSepPunct{\mcitedefaultmidpunct}
{\mcitedefaultendpunct}{\mcitedefaultseppunct}\relax
\EndOfBibitem
\bibitem[Sharma \emph{et~al.}(2020)Sharma, Srinivasan, Sakai, and
  Mitra]{Sharma_2020}
V.~K. Sharma, H.~Srinivasan, V.~G. Sakai and S.~Mitra, \emph{Structural
  Dynamics}, 2020,  051301\relax
\mciteBstWouldAddEndPuncttrue
\mciteSetBstMidEndSepPunct{\mcitedefaultmidpunct}
{\mcitedefaultendpunct}{\mcitedefaultseppunct}\relax
\EndOfBibitem
\bibitem[Metzler \emph{et~al.}(2016)Metzler, Jeon, and Cherstvy]{Metzler_2016}
R.~Metzler, J.-H. Jeon and A.~G. Cherstvy, \emph{Biochimica et Biophysica
  Acta}, 2016, \textbf{1858}, 2451--2467\relax
\mciteBstWouldAddEndPuncttrue
\mciteSetBstMidEndSepPunct{\mcitedefaultmidpunct}
{\mcitedefaultendpunct}{\mcitedefaultseppunct}\relax
\EndOfBibitem
\bibitem[Montroll and Weiss(1965)]{Montroll_1965}
E.~W. Montroll and G.~H. Weiss, \emph{Journal of Mathematical Physics}, 1965,
  \textbf{6}, 167--181\relax
\mciteBstWouldAddEndPuncttrue
\mciteSetBstMidEndSepPunct{\mcitedefaultmidpunct}
{\mcitedefaultendpunct}{\mcitedefaultseppunct}\relax
\EndOfBibitem
\bibitem[Abbott \emph{et~al.}(2003)Abbott, Capper, Davies, Rasheed, and
  Tambyrajah]{Abbott_2003}
A.~P. Abbott, G.~Capper, D.~L. Davies, R.~K. Rasheed and V.~Tambyrajah,
  \emph{Chemical Communications}, 2003, \textbf{1}, 70--71\relax
\mciteBstWouldAddEndPuncttrue
\mciteSetBstMidEndSepPunct{\mcitedefaultmidpunct}
{\mcitedefaultendpunct}{\mcitedefaultseppunct}\relax
\EndOfBibitem
\bibitem[Abbott(2004)]{Abbott_2004}
A.~P. Abbott, \emph{ChemPhysChem}, 2004, \textbf{5}, 1242--1246\relax
\mciteBstWouldAddEndPuncttrue
\mciteSetBstMidEndSepPunct{\mcitedefaultmidpunct}
{\mcitedefaultendpunct}{\mcitedefaultseppunct}\relax
\EndOfBibitem
\bibitem[Smith \emph{et~al.}(2014)Smith, Abbott, and Ryder]{Smith_2014}
E.~L. Smith, A.~P. Abbott and K.~S. Ryder, \emph{Chemical Reviews}, 2014,
  \textbf{114}, 11060--11082\relax
\mciteBstWouldAddEndPuncttrue
\mciteSetBstMidEndSepPunct{\mcitedefaultmidpunct}
{\mcitedefaultendpunct}{\mcitedefaultseppunct}\relax
\EndOfBibitem
\bibitem[Zhang \emph{et~al.}(2012)Zhang, Vigier, Royer, and
  J{\'e}r{\^o}me]{zhang_2012}
Q.~Zhang, K.~D.~O. Vigier, S.~Royer and F.~J{\'e}r{\^o}me, \emph{Chemical
  Society Reviews}, 2012, \textbf{41}, 7108--7146\relax
\mciteBstWouldAddEndPuncttrue
\mciteSetBstMidEndSepPunct{\mcitedefaultmidpunct}
{\mcitedefaultendpunct}{\mcitedefaultseppunct}\relax
\EndOfBibitem
\bibitem[Guchhait \emph{et~al.}(2014)Guchhait, Das, Daschakraborty, and
  Biswas]{Guchhait_2014}
B.~Guchhait, S.~Das, S.~Daschakraborty and R.~Biswas, \emph{The Journal of
  Chemical Physics}, 2014, \textbf{140}, 104514\relax
\mciteBstWouldAddEndPuncttrue
\mciteSetBstMidEndSepPunct{\mcitedefaultmidpunct}
{\mcitedefaultendpunct}{\mcitedefaultseppunct}\relax
\EndOfBibitem
\bibitem[Boisset \emph{et~al.}(2013)Boisset, Menne, Jacquemin, Balducci, and
  Anouti]{Boisset_2013}
A.~Boisset, S.~Menne, J.~Jacquemin, A.~Balducci and M.~Anouti, \emph{Physical
  Chemistry Chemical Physics}, 2013, \textbf{15}, 20054--20063\relax
\mciteBstWouldAddEndPuncttrue
\mciteSetBstMidEndSepPunct{\mcitedefaultmidpunct}
{\mcitedefaultendpunct}{\mcitedefaultseppunct}\relax
\EndOfBibitem
\bibitem[Lin and Sun(1999)]{Lin_1999}
Y.-F. Lin and I.~W. Sun, \emph{Electrochimica Acta}, 1999, \textbf{44},
  2771--2777\relax
\mciteBstWouldAddEndPuncttrue
\mciteSetBstMidEndSepPunct{\mcitedefaultmidpunct}
{\mcitedefaultendpunct}{\mcitedefaultseppunct}\relax
\EndOfBibitem
\bibitem[Smith(2013)]{Smith*_2013}
E.~L. Smith, \emph{Transactions of the IMF}, 2013, \textbf{91}, 241--248\relax
\mciteBstWouldAddEndPuncttrue
\mciteSetBstMidEndSepPunct{\mcitedefaultmidpunct}
{\mcitedefaultendpunct}{\mcitedefaultseppunct}\relax
\EndOfBibitem
\bibitem[Abbott \emph{et~al.}(2012)Abbott, Ttaib, Frisch, Ryder, and
  Weston]{abbott_2012}
A.~P. Abbott, K.~E. Ttaib, G.~Frisch, K.~S. Ryder and D.~Weston, \emph{Physical
  Chemistry Chemical Physics}, 2012, \textbf{14}, 2443--2449\relax
\mciteBstWouldAddEndPuncttrue
\mciteSetBstMidEndSepPunct{\mcitedefaultmidpunct}
{\mcitedefaultendpunct}{\mcitedefaultseppunct}\relax
\EndOfBibitem
\bibitem[Sun \emph{et~al.}(2012)Sun, Zhao, and Zheng]{Sun_2012}
A.~Sun, H.~Zhao and J.~Zheng, \emph{Talanta}, 2012, \textbf{88}, 259--264\relax
\mciteBstWouldAddEndPuncttrue
\mciteSetBstMidEndSepPunct{\mcitedefaultmidpunct}
{\mcitedefaultendpunct}{\mcitedefaultseppunct}\relax
\EndOfBibitem
\bibitem[G{\'o}mez \emph{et~al.}(2011)G{\'o}mez, Cojocaru, Magagnin, and
  Valles]{Gomez_2011}
E.~G{\'o}mez, P.~Cojocaru, L.~Magagnin and E.~Valles, \emph{Journal of
  Electroanalytical Chemistry}, 2011, \textbf{658}, 18--24\relax
\mciteBstWouldAddEndPuncttrue
\mciteSetBstMidEndSepPunct{\mcitedefaultmidpunct}
{\mcitedefaultendpunct}{\mcitedefaultseppunct}\relax
\EndOfBibitem
\bibitem[Chirea \emph{et~al.}(2011)Chirea, Freitas, Vasile, Ghitulica, Pereira,
  and Silva]{Chirea_2011}
M.~Chirea, A.~Freitas, B.~S. Vasile, C.~Ghitulica, C.~M. Pereira and F.~Silva,
  \emph{Langmuir}, 2011, \textbf{27}, 3906--3913\relax
\mciteBstWouldAddEndPuncttrue
\mciteSetBstMidEndSepPunct{\mcitedefaultmidpunct}
{\mcitedefaultendpunct}{\mcitedefaultseppunct}\relax
\EndOfBibitem
\bibitem[Liu \emph{et~al.}(2010)Liu, Yu, Cao, Su, Liu, Zhang, and
  Wang]{Liu_2010}
W.~Liu, Y.~Yu, L.~Cao, G.~Su, X.~Liu, L.~Zhang and Y.~Wang, \emph{Journal of
  hazardous materials}, 2010, \textbf{181}, 1102--1108\relax
\mciteBstWouldAddEndPuncttrue
\mciteSetBstMidEndSepPunct{\mcitedefaultmidpunct}
{\mcitedefaultendpunct}{\mcitedefaultseppunct}\relax
\EndOfBibitem
\bibitem[Guti{\'e}rrez \emph{et~al.}(2011)Guti{\'e}rrez, Carriazo, Tamayo,
  Jim{\'e}nez, Pic{\'o}, Rojo, Ferrer, and {del Monte}]{Gutierrez_2011}
M.~C. Guti{\'e}rrez, D.~Carriazo, A.~Tamayo, R.~Jim{\'e}nez, F.~Pic{\'o}, J.~M.
  Rojo, M.~L. Ferrer and F.~{del Monte}, \emph{Chemistry {\textendash} A
  European Journal}, 2011, \textbf{17}, 10533--10537\relax
\mciteBstWouldAddEndPuncttrue
\mciteSetBstMidEndSepPunct{\mcitedefaultmidpunct}
{\mcitedefaultendpunct}{\mcitedefaultseppunct}\relax
\EndOfBibitem
\bibitem[Morrison \emph{et~al.}(2009)Morrison, Sun, and
  Neervannan]{Morrison_2009}
H.~G. Morrison, C.~C. Sun and S.~Neervannan, \emph{International Journal of
  Pharmaceutics}, 2009, \textbf{378}, 136--139\relax
\mciteBstWouldAddEndPuncttrue
\mciteSetBstMidEndSepPunct{\mcitedefaultmidpunct}
{\mcitedefaultendpunct}{\mcitedefaultseppunct}\relax
\EndOfBibitem
\bibitem[Xie \emph{et~al.}(2016)Xie, Dong, Zhang, Lu, and Ji]{Xie_2016}
Y.~Xie, H.~Dong, S.~Zhang, X.~Lu and X.~Ji, \emph{Green Energy \& Environment},
  2016, \textbf{1}, 195--200\relax
\mciteBstWouldAddEndPuncttrue
\mciteSetBstMidEndSepPunct{\mcitedefaultmidpunct}
{\mcitedefaultendpunct}{\mcitedefaultseppunct}\relax
\EndOfBibitem
\bibitem[Guchhait \emph{et~al.}(2012)Guchhait, Daschakraborty, and
  Biswas]{Guchhait_2012}
B.~Guchhait, S.~Daschakraborty and R.~Biswas, \emph{The Journal of Chemical
  Physics}, 2012, \textbf{136}, 174503\relax
\mciteBstWouldAddEndPuncttrue
\mciteSetBstMidEndSepPunct{\mcitedefaultmidpunct}
{\mcitedefaultendpunct}{\mcitedefaultseppunct}\relax
\EndOfBibitem
\bibitem[Srinivasan \emph{et~al.}(2020)Srinivasan, Sharma, Mukhopadhyay, and
  Mitra]{Srinivasan_2020a}
H.~Srinivasan, V.~K. Sharma, R.~Mukhopadhyay and S.~Mitra, \emph{The Journal of
  Chemical Physics}, 2020, \textbf{153}, 104505\relax
\mciteBstWouldAddEndPuncttrue
\mciteSetBstMidEndSepPunct{\mcitedefaultmidpunct}
{\mcitedefaultendpunct}{\mcitedefaultseppunct}\relax
\EndOfBibitem
\bibitem[Srinivasan \emph{et~al.}(2023)Srinivasan, Sharma, Sakai, Mukhopadhyay,
  and Mitra]{Srinivasan_2023}
H.~Srinivasan, V.~K. Sharma, V.~G. Sakai, R.~Mukhopadhyay and S.~Mitra,
  \emph{The Journal of Physical Chemistry Letters}, 2023, \textbf{14},
  9766--9773\relax
\mciteBstWouldAddEndPuncttrue
\mciteSetBstMidEndSepPunct{\mcitedefaultmidpunct}
{\mcitedefaultendpunct}{\mcitedefaultseppunct}\relax
\EndOfBibitem
\bibitem[Qvist \emph{et~al.}(2011)Qvist, Schober, and Halle]{Qvist_2011}
J.~Qvist, H.~Schober and B.~Halle, \emph{Journal of Chemical Physics}, 2011,
  \textbf{134}, 144508\relax
\mciteBstWouldAddEndPuncttrue
\mciteSetBstMidEndSepPunct{\mcitedefaultmidpunct}
{\mcitedefaultendpunct}{\mcitedefaultseppunct}\relax
\EndOfBibitem
\bibitem[Embs \emph{et~al.}(2013)Embs, Burankova, Reichert, Fossog, and
  Hempelmann]{Embs_2013}
J.~P. Embs, T.~Burankova, E.~Reichert, V.~Fossog and R.~Hempelmann,
  \emph{Journal of the Physical Society of Japan}, 2013, \textbf{82},
  SA003\relax
\mciteBstWouldAddEndPuncttrue
\mciteSetBstMidEndSepPunct{\mcitedefaultmidpunct}
{\mcitedefaultendpunct}{\mcitedefaultseppunct}\relax
\EndOfBibitem
\bibitem[Berrod \emph{et~al.}(2017)Berrod, Ferdeghini, Zanotti, Judeinstein,
  Lairez, Garc{\'i}a~Sakai, Czakkel, Fouquet, and Constantin]{Berrod_2017}
Q.~Berrod, F.~Ferdeghini, J.-M. Zanotti, P.~Judeinstein, D.~Lairez,
  V.~Garc{\'i}a~Sakai, O.~Czakkel, P.~Fouquet and D.~Constantin,
  \emph{Scientific Reports}, 2017, \textbf{7}, 2241\relax
\mciteBstWouldAddEndPuncttrue
\mciteSetBstMidEndSepPunct{\mcitedefaultmidpunct}
{\mcitedefaultendpunct}{\mcitedefaultseppunct}\relax
\EndOfBibitem
\bibitem[Wagle \emph{et~al.}(2015)Wagle, Baker, and Mamontov]{Wagle_2015}
D.~V. Wagle, G.~A. Baker and E.~Mamontov, \emph{The Journal of Physical
  Chemistry Letters}, 2015, \textbf{6}, 2924--2928\relax
\mciteBstWouldAddEndPuncttrue
\mciteSetBstMidEndSepPunct{\mcitedefaultmidpunct}
{\mcitedefaultendpunct}{\mcitedefaultseppunct}\relax
\EndOfBibitem
\bibitem[Mandelbrot and Van~Ness(1968)]{Mandelbrot_1968}
B.~B. Mandelbrot and J.~W. Van~Ness, \emph{SIAM Review}, 1968, \textbf{10},
  422--437\relax
\mciteBstWouldAddEndPuncttrue
\mciteSetBstMidEndSepPunct{\mcitedefaultmidpunct}
{\mcitedefaultendpunct}{\mcitedefaultseppunct}\relax
\EndOfBibitem
\bibitem[Duncan \emph{et~al.}(2000)Duncan, Hu, and {Pasik-Duncan}]{Duncan_2000}
T.~E. Duncan, Y.~Hu and B.~{Pasik-Duncan}, \emph{SIAM Journal on Control and
  Optimization}, 2000, \textbf{38}, 582--612\relax
\mciteBstWouldAddEndPuncttrue
\mciteSetBstMidEndSepPunct{\mcitedefaultmidpunct}
{\mcitedefaultendpunct}{\mcitedefaultseppunct}\relax
\EndOfBibitem
\bibitem[Kou and Xie(2004)]{Kou_2004}
S.~C. Kou and X.~S. Xie, \emph{Physical Review Letters}, 2004, \textbf{93},
  180603\relax
\mciteBstWouldAddEndPuncttrue
\mciteSetBstMidEndSepPunct{\mcitedefaultmidpunct}
{\mcitedefaultendpunct}{\mcitedefaultseppunct}\relax
\EndOfBibitem
\bibitem[Srinivasan \emph{et~al.}(2024)Srinivasan, Sharma, Garc\'{\i}a~Sakai,
  and Mitra]{Srinivasan_2024}
H.~Srinivasan, V.~K. Sharma, V.~Garc\'{\i}a~Sakai and S.~Mitra, \emph{Phys.
  Rev. Lett.}, 2024, \textbf{132}, 058202\relax
\mciteBstWouldAddEndPuncttrue
\mciteSetBstMidEndSepPunct{\mcitedefaultmidpunct}
{\mcitedefaultendpunct}{\mcitedefaultseppunct}\relax
\EndOfBibitem
\bibitem[Busselez \emph{et~al.}(2011)Busselez, Lefort, Ghoufi, Beuneu, Frick,
  Affouard, and Morineau]{Busselez_2011}
R.~Busselez, R.~Lefort, A.~Ghoufi, B.~Beuneu, B.~Frick, F.~Affouard and
  D.~Morineau, \emph{Journal of Physics: Condensed Matter}, 2011, \textbf{23},
  505102\relax
\mciteBstWouldAddEndPuncttrue
\mciteSetBstMidEndSepPunct{\mcitedefaultmidpunct}
{\mcitedefaultendpunct}{\mcitedefaultseppunct}\relax
\EndOfBibitem
\bibitem[Kofu \emph{et~al.}(2018)Kofu, Faraone, Tyagi, Nagao, and
  Yamamuro]{Kofu_2018}
M.~Kofu, A.~Faraone, M.~Tyagi, M.~Nagao and O.~Yamamuro, \emph{Physical Review
  E}, 2018, \textbf{98}, 042601\relax
\mciteBstWouldAddEndPuncttrue
\mciteSetBstMidEndSepPunct{\mcitedefaultmidpunct}
{\mcitedefaultendpunct}{\mcitedefaultseppunct}\relax
\EndOfBibitem
\bibitem[Arbe \emph{et~al.}(2003)Arbe, Colmenero, Alvarez, Monkenbusch,
  Richter, Farago, and Frick]{Arbe_2003}
A.~Arbe, J.~Colmenero, F.~Alvarez, M.~Monkenbusch, D.~Richter, B.~Farago and
  B.~Frick, \emph{Physical Review E}, 2003, \textbf{67}, 051802\relax
\mciteBstWouldAddEndPuncttrue
\mciteSetBstMidEndSepPunct{\mcitedefaultmidpunct}
{\mcitedefaultendpunct}{\mcitedefaultseppunct}\relax
\EndOfBibitem
\bibitem[Capponi \emph{et~al.}(2009)Capponi, Arbe, Alvarez, Colmenero, Frick,
  and Embs]{Capponi_2009}
S.~Capponi, A.~Arbe, F.~Alvarez, J.~Colmenero, B.~Frick and J.~P. Embs,
  \emph{The Journal of Chemical Physics}, 2009, \textbf{131}, 204901\relax
\mciteBstWouldAddEndPuncttrue
\mciteSetBstMidEndSepPunct{\mcitedefaultmidpunct}
{\mcitedefaultendpunct}{\mcitedefaultseppunct}\relax
\EndOfBibitem
\bibitem[Colmenero \emph{et~al.}(2002)Colmenero, Alvarez, and
  Arbe]{Colmenero_2002}
J.~Colmenero, F.~Alvarez and A.~Arbe, \emph{Physical Review E}, 2002,
  \textbf{65}, 041804\relax
\mciteBstWouldAddEndPuncttrue
\mciteSetBstMidEndSepPunct{\mcitedefaultmidpunct}
{\mcitedefaultendpunct}{\mcitedefaultseppunct}\relax
\EndOfBibitem
\bibitem[Wang \emph{et~al.}(2009)Wang, Anthony, Bae, and Granick]{Wang_2009}
B.~Wang, S.~M. Anthony, S.~C. Bae and S.~Granick, \emph{Proceedings of the
  National Academy of Sciences}, 2009, \textbf{106}, 15160--15164\relax
\mciteBstWouldAddEndPuncttrue
\mciteSetBstMidEndSepPunct{\mcitedefaultmidpunct}
{\mcitedefaultendpunct}{\mcitedefaultseppunct}\relax
\EndOfBibitem
\bibitem[Chubynsky and Slater(2014)]{Chubynsky_2014}
M.~V. Chubynsky and G.~W. Slater, \emph{Physical Review Letters}, 2014,
  \textbf{113}, 098302\relax
\mciteBstWouldAddEndPuncttrue
\mciteSetBstMidEndSepPunct{\mcitedefaultmidpunct}
{\mcitedefaultendpunct}{\mcitedefaultseppunct}\relax
\EndOfBibitem
\bibitem[Chaudhuri \emph{et~al.}(2007)Chaudhuri, Berthier, and
  Kob]{Chaudhuri_2007}
P.~Chaudhuri, L.~Berthier and W.~Kob, \emph{Physical Review Letters}, 2007,
  \textbf{99}, 060604\relax
\mciteBstWouldAddEndPuncttrue
\mciteSetBstMidEndSepPunct{\mcitedefaultmidpunct}
{\mcitedefaultendpunct}{\mcitedefaultseppunct}\relax
\EndOfBibitem
\bibitem[Rusciano \emph{et~al.}(2022)Rusciano, Pastore, and
  Greco]{Rusciano_2022}
F.~Rusciano, R.~Pastore and F.~Greco, \emph{Physical Review Letters}, 2022,
  \textbf{128}, 168001\relax
\mciteBstWouldAddEndPuncttrue
\mciteSetBstMidEndSepPunct{\mcitedefaultmidpunct}
{\mcitedefaultendpunct}{\mcitedefaultseppunct}\relax
\EndOfBibitem
\bibitem[Pastore \emph{et~al.}(2021)Pastore, Ciarlo, Pesce, Greco, and
  Sasso]{Pastore_2021}
R.~Pastore, A.~Ciarlo, G.~Pesce, F.~Greco and A.~Sasso, \emph{Physical Review
  Letters}, 2021, \textbf{126}, 158003\relax
\mciteBstWouldAddEndPuncttrue
\mciteSetBstMidEndSepPunct{\mcitedefaultmidpunct}
{\mcitedefaultendpunct}{\mcitedefaultseppunct}\relax
\EndOfBibitem
\bibitem[Berthier \emph{et~al.}(2007)Berthier, Biroli, Bouchaud, Kob, Miyazaki,
  and Reichman]{Berthier_2007}
L.~Berthier, G.~Biroli, J.~P. Bouchaud, W.~Kob, K.~Miyazaki and D.~R. Reichman,
  \emph{The Journal of Chemical Physics}, 2007, \textbf{126}, 184504\relax
\mciteBstWouldAddEndPuncttrue
\mciteSetBstMidEndSepPunct{\mcitedefaultmidpunct}
{\mcitedefaultendpunct}{\mcitedefaultseppunct}\relax
\EndOfBibitem
\bibitem[Berthier and Kob(2007)]{Berthier_2007a}
L.~Berthier and W.~Kob, \emph{Journal of Physics: Condensed Matter}, 2007,
  \textbf{19}, 205130\relax
\mciteBstWouldAddEndPuncttrue
\mciteSetBstMidEndSepPunct{\mcitedefaultmidpunct}
{\mcitedefaultendpunct}{\mcitedefaultseppunct}\relax
\EndOfBibitem
\bibitem[Barkai and Burov(2020)]{Barkai_2020}
E.~Barkai and S.~Burov, \emph{Physical Review Letters}, 2020, \textbf{124},
  060603\relax
\mciteBstWouldAddEndPuncttrue
\mciteSetBstMidEndSepPunct{\mcitedefaultmidpunct}
{\mcitedefaultendpunct}{\mcitedefaultseppunct}\relax
\EndOfBibitem
\bibitem[Crupi \emph{et~al.}(2003)Crupi, Majolino, Migliardo, and
  Venuti]{Crupi_2003}
V.~Crupi, D.~Majolino, P.~Migliardo and V.~Venuti, \emph{The Journal of
  Chemical Physics}, 2003, \textbf{118}, 5971--5978\relax
\mciteBstWouldAddEndPuncttrue
\mciteSetBstMidEndSepPunct{\mcitedefaultmidpunct}
{\mcitedefaultendpunct}{\mcitedefaultseppunct}\relax
\EndOfBibitem
\bibitem[Sobolev \emph{et~al.}(2007)Sobolev, Novikov, and Pieper]{Sobolev_2007}
O.~Sobolev, A.~Novikov and J.~Pieper, \emph{Chemical Physics}, 2007,
  \textbf{334}, 36--44\relax
\mciteBstWouldAddEndPuncttrue
\mciteSetBstMidEndSepPunct{\mcitedefaultmidpunct}
{\mcitedefaultendpunct}{\mcitedefaultseppunct}\relax
\EndOfBibitem
\bibitem[Banerjee \emph{et~al.}(2020)Banerjee, Ghorai, Das, Rajbangshi, and
  Biswas]{Banerjee_2020}
S.~Banerjee, P.~K. Ghorai, S.~Das, J.~Rajbangshi and R.~Biswas, \emph{The
  Journal of Chemical Physics}, 2020, \textbf{153}, 234502\relax
\mciteBstWouldAddEndPuncttrue
\mciteSetBstMidEndSepPunct{\mcitedefaultmidpunct}
{\mcitedefaultendpunct}{\mcitedefaultseppunct}\relax
\EndOfBibitem
\end{mcitethebibliography}
\bibliographystyle{rsc} 

\end{document}